\documentclass[12pt,letterpaper]{article}
\pdfoutput=1
\usepackage{amsmath}
\usepackage{amsfonts}
\usepackage{amssymb}
\usepackage{graphicx}

\usepackage{graphicx}
\usepackage{epstopdf}
\usepackage{amsmath}
\usepackage{amsfonts}
\usepackage{amssymb}
\usepackage{color}

\newcommand{\be}{\begin{equation}}
\newcommand{\ee}{\end{equation}}
\newcommand{\bea}{\begin{eqnarray}}
\newcommand{\eea}{\end{eqnarray}}
\newcommand{\barr}{\begin{array}}
\newcommand{\earr}{\end{array}}

\def\beq{\begin{equation}}
\def\eeq{\end{equation}}
\def\be{\begin{equation}}
\def\ee{\end{equation}}
\def\bea{\begin{eqnarray}}
\def\eea{\end{eqnarray}}
\def\d{{\partial}}
\def\fnlequil{f_{NL}^{\rm equil.}}
\def\fnlloc{f_{NL}^{\rm loc.}}

\def\fnlorthog{f_{NL}^{\rm orthog.}}

\def\hfnlequil{{\hat f}_{NL}^{\rm equil.}}
\def\hfnlorthog{{\hat f}_{NL}^{\rm orthog.}}
\def\mpl{M_{\rm Pl}}
\def\nn{\nonumber}
\def\n{{\bf \widehat n}}

\begin{document}

\vspace{5mm}
\vspace{0.5cm}
\begin{center}

\def\thefootnote{\fnsymbol{footnote}}

{\Large \bf Non-Gaussianities in Single Field Inflation\\[0.1cm] 
and their Optimal Limits  \\[0.4cm]
from the WMAP 5-year Data}
\\[0.7cm]
{\large Leonardo Senatore$^{a}$, Kendrick M. Smith$^{b}$ and Matias Zaldarriaga$^{c}$}
\\[0.7cm]
{\normalsize { \sl $^{a}$ School of Natural Sciences, Institute for Advanced Study, \\Olden Lane, Princeton, NJ 08540, USA}}\\
\vspace{.3cm}
{\normalsize { \sl $^{b}$ Institute of Astronomy,\\
Cambridge University,  Cambridge, CB3 0HA, UK}}\\

\vspace{.3cm}
{\normalsize { \sl $^{c}$ Jefferson Physical Laboratory and Center for Astrophysics, Harvard University, Cambridge, MA 02138, USA}}\\

\end{center}

\vspace{.8cm}

\hrule \vspace{0.3cm}
{\small  \noindent \textbf{Abstract} \\[0.3cm]
\noindent
Using the recently developed effective field theory of inflation, we argue that the size and the shape of the non-Gaussianities generated by single-field inflation are generically well described by two parameters: $\fnlequil$, which characterizes the size of the signal that is peaked on equilateral configurations, and  $\fnlorthog$, which instead characterizes the size of the signal which is peaked both on equilateral configurations and flat-triangle configurations (with opposite signs). The shape of non-Gaussianities associated with $\fnlorthog$ is orthogonal to the one associated to $\fnlequil$, and former analysis have been mostly blind to it.  We perform the optimal analysis of the WMAP 5-year data for both of these parameters. We find no evidence of non-Gaussianity, and we have the following constraints: $-125\le\fnlequil\le435$, $-369\le \fnlorthog\le 71$ at 95\% CL. We show that both of these constraints can be translated into limits on parameters of the Lagrangian of single-field inflation. For one of them, the speed of sound of the inflaton fluctuations, we find that it is either bounded to be $c_s\ge 0.011$ at 95\% CL. or alternatively to be so small that the higher-derivative kinetic term dominate at horizon crossing. We are able to put similar constraints on the other operators of the inflaton Lagrangian.}
 \vspace{0.5cm}
\hrule



\newpage

\section{Introduction}

The interest in measuring a possible non-Gaussianity of the primordial density perturbation has grown in the last few years due mainly to two reasons:  on the one hand, new experiments have finally become sensitive enough to measure even relatively small deviations from  Gaussian statistics; on the other hand, new theoretical insights have shown that while standard slow-roll inflation where the inflaton has a canonical kinetic term predicts an undetectable level of non-Gaussianity  \cite{Maldacena:2002vr}, many other inflationary models  can give rise to a level of non-Gaussianity already detectable  by current experiments. Furthermore it has been shown \cite{Babich:2004gb,Cheung:2007st} that a detection of  non-Gaussianity would carry an unprecedented amount of information on the dynamics that drove inflation, allowing us to explore the interaction Lagrangian of the inflaton.

The current constraints on the level of non-Gaussianity of the primordial density perturbations already imply that the deviation from a Gaussian spectrum is at most at the percent level. This means that the most natural observables to look at are ones which are zero in the presence of an exactly Gaussian spectrum. This is the case for the bispectrum
\be
\langle \Phi_{\vec k_1} \Phi_{\vec  k_2} \Phi_{\vec k_3} \rangle=(2\pi)^3 \delta^{(3)}( \sum_i \vec{k}_i ) F(k_1,k_2,k_3)\ ,
\ee
where $\Phi$ is the Newtonian potential. 
Because of translation invariance, the sum of the spatial $k$'s must form a closed triangle. In the case of the non-Gaussianities produced by inflation, approximate scale invariance implies that, under rescaling of the momenta, the $F$ above scales approximately as $1/k^6$, where $k$ is the typical size of one of the momenta. Finally, rotation invariance forces $F$ to be only a function of the shape of the triangle (which one can think as being parameterized by the ratios of the two lowest $k$'s with respect to the highest $k$ in the triangle). In some models $F$ can deviate from scale invariance in a way that might have some measurable effects if one looks at non-Gaussianities on very different scales \cite{LoVerde:2007ri}. At present this effect is  not very important for Cosmic Microwave Background (CMB) experiments given the lack of a detection. Therefore we will neglect this in what follows.

So far, all analyses both in the CMB and in Large Scale Structure (LSS) have concentrated just on two kinds of non-Gaussianities. The first one is the so called {\it local} type, in which  $\Phi$ is given by \cite{Komatsu:2001rj}
\be
\Phi(\vec x)=\Phi^{(g)}(\vec x)+\fnlloc \left(\Phi^{(g)}(\vec x)^2-\langle\Phi^{(g)}(\vec x)^2\rangle\right)\ ,
\ee
where $\Phi^{(g)}$ is a Gaussian stochastic variable. $\fnlloc$ determines the size of the non-Gaussianities, and is the parameter being constrained. In momentum space this kind of non-Gaussianity implies that the signal is peaked on `squeezed' triangles where one of the sides is much smaller than the other two. This kind of non-Gaussianity is predicted in some multi-field inflationary models, see for example \cite{Lyth:2002my,Zaldarriaga:2003my}, or in the recently proposed new bouncing cosmology \cite{Creminelli:2007aq}. There is  a theorem that states that the non-Gaussian signal in the squeezed limit that can be produced by single-field inflation, has to be proportional to the tilt of the power spectrum \cite{Maldacena:2002vr,Creminelli:2004yq,Cheung:2007sv}. This means that, unless a strong deviation from scale invariance in the power spectrum is found, detection of such a non-Gaussian signal would definitively rule out all single-field inflationary models. The current constraint on this kind of signal from the WMAP 5yr data is given by \cite{Smith:2009jr}:
\be
-4\le\fnlloc\le 80\quad {\rm at \ 95\%\ \ CL}\ ,
\ee
where the optimal estimator found in \cite{Creminelli:2005hu}  was implemented with the pipeline developed in  \cite{Smith:2006ud}.
When combined with the constraint from LSS \cite{Slosar:2008hx}, this gives $-1\le\fnlloc\le63$ at 95\% CL.

The second kind of non-Gaussianity that has been constrained is the so-called {\it equilateral} kind. In this case, the signal is peaked on equilateral-triangle configurations in Fourier space. This is the shape that is produced in some proposed single-field or multi-field inflationary models, such as DBI inflation \cite{Alishahiha:2004eh,Langlois:2008wt,Arroja:2008yy}, or Ghost inflation \cite{ArkaniHamed:2003uz,Senatore:2004rj}, where the inflaton has large derivative interactions, or  Trapped inflation \cite{Green:2009ds}, where production of massive particles slows down the inflaton. This shape is quite different from the local shape, and requires a dedicated analysis. The current best limits are given by the WMAP team, who applied the technique developed in  \cite{Creminelli:2005hu} to the WMAP 5yr data:  $-151\le\fnlequil\le 253$ at 95\% CL \cite{Komatsu:2008hk}. In this paper we will improve the above limit by performing the optimal analysis for the same parameter $\fnlequil$, using the estimator first presented in \cite{Creminelli:2005hu}, and implementing it with the improved pipeline of \cite{Smith:2006ud}. Furthermore, this allows us also to marginalize over the amplitude of the templates of the galactic foregrounds, performing an improved treatment of the foreground with respect to the cleaning procedure applied in \cite{Komatsu:2008hk}. After implementing the optimal analysis, we find no evidence of non-Gaussianity:
\be
-125\le \fnlequil\le 435  \quad {\rm at \ 95\%\ \ CL}\ ,
\ee
This constraint is obtained by assuming that other forms of primordial non-Gaussianities are absent.
Though this limit is optimal, it is a little bit milder than the one claimed by the WMAP collaboration on the same set of data. As we explain in detail in sec.~\ref{sec:constraints_error}, we believe that previous analyses have
underestimated the error $\sigma(\fnlequil$)
\footnote{We acknowledge Eiichiro Komatsu for collaboration regarding some comparisons of our code with the one used by the WMAP team.}.

Improving the limit on $\fnlequil$ is not however the main purpose of the present paper. Thanks to the development of the effective field theory of inflation \cite{Cheung:2007st}, we will be able to argue that the parameter space of the non-Gaussianities produced by the most general single-field models, where the inflaton fluctuations have an approximate shift symmetry, is in reality larger than the one characterized by $\fnlequil$. It consists of any linear combination of two independent shapes: the equilateral one, and a new one that we call {\it orthogonal}. This orthogonal shape is peaked both on equilateral-triangle configurations and on flattened-triangle configurations (where the two lowest-$k$ sides are equal exactly to half of the highest-$k$ side). The sign in this two limits is opposite.  The new shape is orthogonal to the equilateral one (we define  a scalar product  later in the text). We define $\fnlorthog$ to parameterize the amplitude of the orthogonal template,  and we then argue that former analysis of the bispectrum for the local and the equilateral shape have been quite insensitive to the orthogonal shape which therefore could be rather large in the data and still  not have been detected. After performing the optimal analysis, we find no evidence of a non-zero $\fnlorthog$, and we obtain the constraint:
\be
-369\le \fnlorthog\le 71 \quad {\rm at \ 95\%\ \ CL}\ .
\ee 

In addition, thanks again to the effective field theory formalism, we are able to map the constraint we find on $\fnlequil$ and $\fnlorthog$ into constraints on the coefficients of  some  operators of the Lagrangian for the inflationary fluctuations  (under of course the assumption that there is only one relevant light degree of freedom during the inflationary phase). We believe that being able to put constraints or potentially measure the coefficients of the interaction Lagrangian in inflation is quite a remarkable fact, given that the energy scale at which this can happen could be huge. For the readers familiar with particle physics and collider physics, loosely speaking the limits we will put on the inflaton Lagrangian can be seen as the analogue of the constraints on the higher dimension operators that come from precision electroweak tests \cite{Peskin:1990zt,Peskin:1991sw,Barbieri:2004qk}. The only difference is that while in collider physics we might hope to produce some of the particles mediating new processes on shell at particle accelerators, it is very unlikely that we will be ever able to produce directly the inflaton particle. We believe this makes the formalism we are developing even more important. This comparison with collider physics is somewhat natural, as having non-Gaussianities in the sky, with their huge amount of information associated to the different triangular configurations, is very similar to having in an accelerator a cross section as a function of the angle of the outgoing particles.  The inclusion of the orthogonal shape in the analysis allows us to put constraints on the parameter space of the interactions of the inflationary fluctuations. Since the presence of  a particular kind of non-Gaussianity is associated to a small speed of sound $c_s$ of the fluctuations, we can constrain $c_s$ to be larger than:
\be
c_s\ge 0.011 \qquad {\rm at \ 95\%\ \ CL}\ .
\ee
or to be smaller than
\be
c_s\lesssim 10^{-2}\;(d_2+d_3)^{2/5}\ ,
\ee
In this last case the higher-derivative kinetic term, whose size is proportional to $(d_2+d_3)$, is important at horizon crossing and the non-Gaussianities depend on other coefficients. 
As we argue in the text, without the inclusion of the orthogonal shape into the analysis, it would have been impossible to set this bound without rather strong assumptions.
We are then able to derive similar constraints on other three parameters of the inflaton Lagrangian. In some region of the parameter space, consistent inflationary models have a negative squared speed of sound $c_s^2$ for the fluctuations at horizon crossing. This leads to an exponential growth of the perturbations before horizon crossing. The analysis of the WMAP data shows that these models are practically ruled out at 95\% CL.

The paper is organized as follows. In sec.~\ref{sec:EFT} we briefly review the effective field theory of inflation, and we show that there are two independent shapes that need to be analyzed. In sec.~\ref{sec:strategy} we show how we can analyze the two shapes in the CMB data. In sec.~\ref{sec:analysis}, we actually give the results of the analysis of the WMAP data and the constraints on the Lagrangian for the inflationary fluctuations. In sec.~\ref{sec:summary}, we summarize our results.

\section{Effective Field Theory of Inflation and Shape of non-Gaussianities\label{sec:EFT}}

An effective field theory description for the fluctuations of the inflaton has been developed in~\cite{Cheung:2007st}. By unifying in one description all single-field models, it allows us to explore in full generality the signature space of single-field inflation, in its broadest sense. In inflation there is a physical clock that controls when inflation ends. This means that time translations are spontaneously broken, and that therefore there is a Goldstone boson associated with this symmetry breaking. As usual, the Lagrangian of this Goldstone boson is highly constrained by the symmetries of the problem, in this case the fact that the spacetime is approximately de Sitter space, in the sense that $|\dot H|/H^2\ll1$. The Goldstone boson, that we can call $\pi$, can be thought of as being equivalent, in standard models of inflation driven by a scalar field, to the perturbations in the scalar field $\delta\phi$. The relation valid at linear order is $\pi=\delta\phi/\dot\phi$, where $\dot\phi$ is the speed of the background solution. We stress that the description in terms of the Goldstone boson $\pi$  is more general than this and it does not assume the presence of a fundamental scalar field. Although here and in the rest of the paper we often refer to the Lagrangian for the Goldstone boson as the Lagrangian for single-field inflation, this should be meant in the broadest sense that there is only one light relevant degree of freedom during the inflationary phase. It does not mean that the background solution is generated by a fundamental scalar field: the Lagrangian for the inflationary fluctuations expressed in terms of the Goldstone boson is universal and independent of the details through which the background solution is generated.

The Goldstone boson $\pi$ is related to the standard curvature perturbation $\zeta$ by the relation $\zeta=-H\pi$, which is valid at linear order and at leading order in the generalized slow roll parameters.
The most general Lagrangian for the Goldstone boson is given by \cite{Cheung:2007st,Cheung:2007sv}:
\bea\label{eq:Lagrangian}
 S_\pi=\int d^4 x \sqrt{-g}&&\!\!\!\!\!\!\!\!\! \left[-\frac{\mpl^2\dot H}{c_s^2}\left(\dot\pi^2-c_s^2\frac{1}{a^2}(\partial_i\pi)^2\right)
\right.\\ \nonumber &&\!\!\!\!\!\!\!+\left.\frac{\dot H \mpl^2}{c_s^2}(1-c_s^2)\dot\pi\frac{1}{a^2}(\partial_i\pi)^2-\frac{\dot H \mpl^2}{c_s^2}(1-c_s^2)\left(1+\frac{2}{3}\frac{\tilde c_3}{c_s^2}\right)\dot\pi^3\right.\\
&&\!\!\!\!\!\!\! \left.-\frac{d_1}{4} H M^3\left(6\,\dot\pi^2+\frac{1}{a^2}(\d_i\pi)^2\right)
-\frac{(d_2+d_3)}{2}M^2 \frac{1}{a^4}(\d_i^2\pi)^2-\frac{1}{4}d_1 M^3 \frac{1}{a^4}(\d_j^2\pi) (\d_i\pi)^2\right.\nonumber\\
&&\!\!\!\!\!\!\!+\ \ldots\ \Big]\ , \nonumber
\eea
where here we have assumed that the Goldstone boson is protected by an approximate shift symmetry that allows us to neglect terms where $\pi$ appears without a derivative acting on it. Let us explain the symbols that appear in the above action. For the large level of non-Gaussianities we will be interested in we can neglect the metric perturbations (see \cite{Cheung:2007st,Cheung:2007sv} for a discussion about this approximation) and we can take the metric $g_{\mu\nu}$ to be unperturbed FRW:
\begin{equation}
ds^2=-dt^2+a(t)^2d\vec x^2 \ .
\end{equation}
$H$ is the standard Hubble parameter, $c_s$ is the speed of sound of the fluctuations, which, since the background is not Lorentz invariant, does not need to be equal to one. $M$ is a free parameter with dimension of mass, while $d_1,\, d_2,\,d_3$ and $\tilde c_3$ are  dimensionless parameters that as we will explain later are expected to be of order one~\footnote{Notice that $\tilde c_3=c_s^2 c_3$, where $c_3$ is the parameter defined in \cite{Cheung:2007sv}. The notation is chosen to be as close as possible to the one of \cite{Cheung:2007sv}.}. The dots represent higher derivative terms that in general give a negligible contribution to the observables, or quartic or higher order terms that we neglect because here we are interested in the 3-point function.

The third line of (\ref{eq:Lagrangian}) contains one time kinetic term and two spatial kinetic terms: the first two are standard ones whose coefficient is proportional to $H$, and the other one is a higher derivative term. There is also a higher-derivative cubic term. These terms become important when inflation happens very close to de Sitter space, in the sense that  \cite{Cheung:2007st}
\be \label{eq:ds_limit}
\mpl^2|\dot H|\lesssim {\rm Max}(\;d_1 H M^3 / 4,\,(d_2+d_3) M^2 H^2/(2\, c_s^2)\;)\ .
\ee
This is so because the coefficient of $(\d_i\pi)^2$ in the first line (and in some limits also the one of $\dot\pi^2$) is forced by the symmetries of the problem to be proportional to $\dot H\mpl^2$. As typical in the effective field theory approach, this fact appears so directly in the Lagrangian because the symmetries are fully exploited in its construction.  In the limit in which the system is very close to de Sitter, this term can become subleading with respect to the ones of the third line of eq.~(\ref{eq:Lagrangian}), which are either a higher-derivative term, or a standard kinetic term, whose coefficient is however proportional to $H$. This happens when 
(\ref{eq:ds_limit}) is satisfied, in which case the system approaches Ghost Inflation \cite{Cheung:2007st,Cheung:2007sv,ArkaniHamed:2003uz,Senatore:2004rj} or its generalization \cite{Cheung:2007st,Cheung:2007sv,Creminelli:2006xe}. We will therefore neglect these three terms for the moment and we will come back to them later~\footnote{Strictly speaking, $c_s$ is the speed of sound of the fluctuations only in the limit in which we neglect the term proportional to $d_1$, which is a good approximation when (\ref{eq:ds_limit}) is not satisfied.}.


We notice that, away from the near-de-Sitter limit, the interaction Lagrangian for the Goldstone boson contains, at leading order in derivatives, just two interaction operators: $\dot\pi (\d_i\pi)^2$ and $\dot\pi^3$. Notice that their coefficients are not fixed by the constraint of being close to a de Sitter phase; they can be large, and therefore induce large and detectable non-Gaussianities. These coefficients however are not completely free. The coefficient of $\dot\pi(\d_i\pi)^2$ is fixed to be correlated with the speed of sound of the fluctuations (again, this appears so evidently, thanks to the effective field theory approach). The coefficient becomes large in the limit of small speed of sound, which explains why a small speed of sound is associated to large non-Gaussianities. The coefficient of $\dot\pi^3$ is instead dependent also on $\tilde c_3$. We expect this number to be order one.  This is the case for example in the UV complete model of DBI Inflation \cite{Alishahiha:2004eh}, where $\tilde c_3=3(1-c_s^2)/2$ and  $c_s\ll 1$. From the effective field theory point of view, when $\tilde c_3$ is of order one, the  strong coupling scale induced by the operator $\dot\pi^3$ is the same as the one induced by the operator $\dot\pi(\d_i\pi)^2\ $. The unitarity cutoff induced by $\dot\pi(\partial_i\pi)^2$ is in fact given by:
\be
\Lambda^4_{\dot\pi(\d_i\pi)^2}\sim16\pi^2 \mpl^2|\dot H| \frac{c_s^5}{(1-c_s^2)^2}\ ,
\ee
while the one due to $\dot\pi^3$ is given by
\be
\Lambda^4_{\dot\pi^3}\sim\Lambda^4_{\dot\pi(\d_i\pi)^2}\cdot \frac{1}{\left(c_s^2+2\tilde c_3/3\right)^2}\ ,
\ee
which are indeed of the same order for $\tilde c_3$ of order one. It is also easy to estimate that if $\tilde c_3$ is order one, loop corrections renormalize the coefficients of the two operators only at order one level. This means that $\tilde c_3$ of order one, and a small speed of sound $c_s$, are technically natural~\footnote{This discussion has assumed that the operator proportional to $(d_2+d_3)$ is negligible up the cutoff. This is the case only if $(d_2+d_3)\lesssim c_s$. When this inequality is violated, $(d_2+d_3)$ must be positive and, as we will later explain more in detail, the natural value of $\tilde c_3$ gets scaled down by a factor of order $c_s^{1/4}/(d_2+d_3)^{1/4}$, which is clearly a negligible correction for $(d_2+d_3)$ of order one and for the values of $c_s$ that are currently allowed by the data.}.

The three point function of the Newtonian potential $\Phi$ has the usual form
\be
\langle \Phi_{\vec k_1} \Phi_{\vec k_2} \Phi_{\vec k_3} \rangle=(2\pi)^3 \delta^{(3)}( \sum_i \vec{k}_i ) F(k_1,k_2,k_3)\ .
\ee
Here 
\be
\Phi=\frac{3}{5}\zeta\ ,
\ee
where  $\zeta$ is the curvature perturbation of comoving slices. This relationship is valid, at first order, out of the horizon, during matter domination.
The $\delta-$function comes from translation invariance and it tells us that the 3-point function is a function of closed triangles in momentum space. 
For single-field inflation, $F$ can be read off from:
\be
\langle \Phi_{\vec k_1} \Phi_{\vec k_2} \Phi_{\vec k_3} \rangle =
  -\left(\frac{3}{5}\right)^3 H^3 \langle \pi_{\vec k_1} \pi_{\vec k_2} \pi_{\vec k_3} \rangle  
  =(2\pi)^3 \delta^{(3)}( \sum_i \vec{k}_i )\left(F_{\dot\pi(\d_i\pi)^2}(k_1,k_2,k_3)+F_{\dot\pi^3}(k_1,k_2,k_3)\right)\
  \ee
where $F_{\dot\pi(\d_i\pi)^2}$ is the shape generated by the operator $\dot\pi(\d_i\pi)^2$, and $F_{\dot\pi^3}$ is instead the one generated by the operator $\dot\pi^3$. In the limit in which we are far enough from de Sitter (in the sense of the inequality (\ref{eq:ds_limit})), and in which we consider an approximate shift symmetry for the Goldstone boson, the resulting form of the non-Gaussianity is given by \cite{Cheung:2007sv}~\footnote{In a different  and somewhat less general formalism, this expression was obtained also in \cite{Chen:2006nt}, where it was already pointed out that the models considered in  \cite{Chen:2006nt} admitted two independent shapes for the non-Gaussianities.}:
\bea\label{eq:shapezero}
&&F_{\dot\pi(\d_i\pi)^2}(k_1,k_2,k_3)=-\frac{5}{12}\left(1-\frac{1}{c_s^2}\right)\cdot\Delta_\Phi^2
 \\ \nonumber
&&\quad\qquad\times\frac{\left(
24 K_3{}^6- 8 K_2{}^2 K_3{}^3 K_1- 8 K_2{}^4 K_1{}^2+22 K_3{}^3 K_1{}^3- 6 K_2{}^2 K_1{}^4+ 2 K_1{}^6\right)}{K_3{}^9K_1{}^3}\ , \\ \nonumber
&&F_{\dot\pi^3}(k_1,k_2,k_3)=\frac{20}{3}\left(1-\frac{1}{c_s^2}\right)\,(\tilde c_3+\frac{3}{2}c_s^2)\,\cdot\Delta_\Phi^2\cdot\frac{1}{K_3{}^3K_1{}^3}\ .
\eea 
Here we have used that
\be
\langle\Phi(\vec k_1)\Phi(\vec k_2)\rangle=(2\pi)^3\delta^{(3)}(\vec k_1+\vec k_2)\frac{\Delta_\Phi}{k^3}\ ,
\ee
with 
\be
\Delta_\Phi=\frac{9}{25}\frac{H^2}{4\,\epsilon\,  c_s\, \mpl^2}\ ,
\ee
and $\epsilon=-\dot H/H^2$.
We have also defined 
\begin{eqnarray}
K_1 &=& k_1 + k_2 + k_3\ , \\ \nn
K_2 &=& \left( k_1 k_2 + k_2 k_3 + k_3 k_1 \right)^{1/2}\ , \\ \nn
K_3 &=& \left( k_1 k_2 k_3 \right)^{1/3}\ .
\end{eqnarray}

\begin{figure}[htp]
\centering
\includegraphics[width=8.2cm]{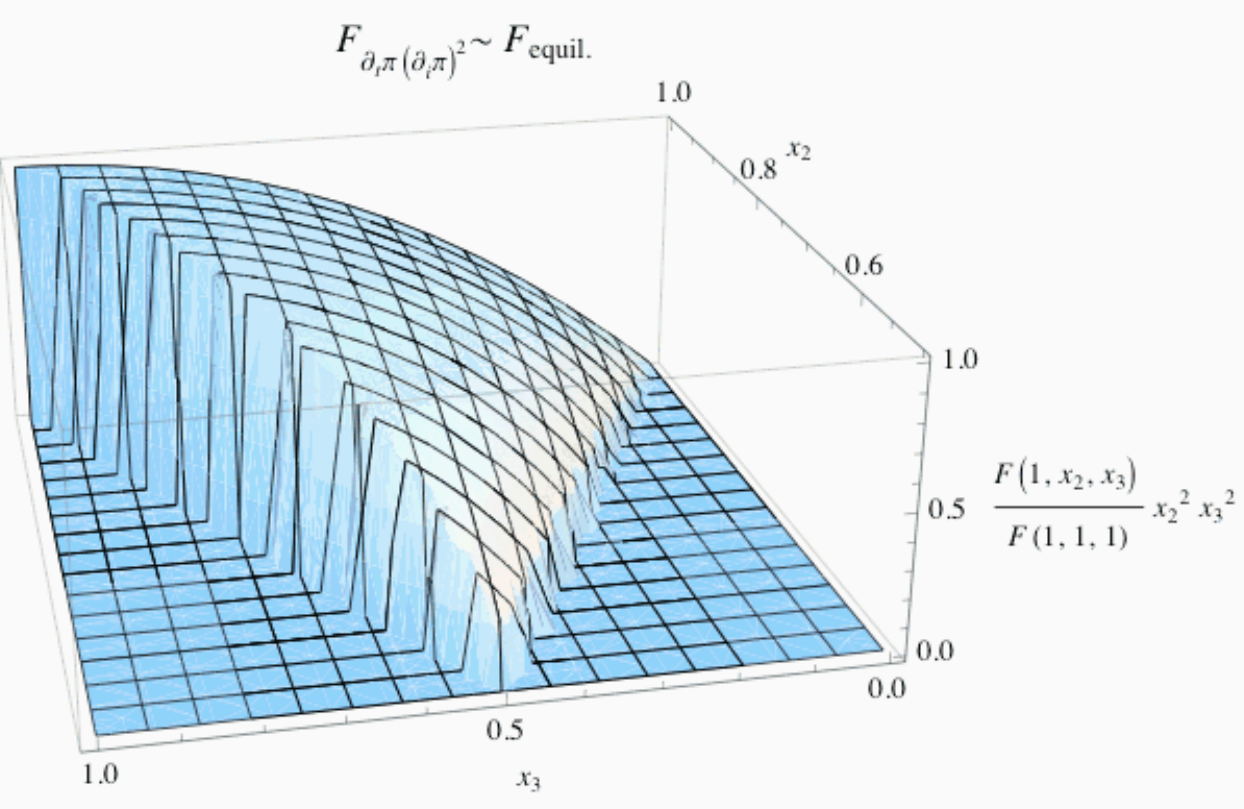}
\includegraphics[width=8.2cm]{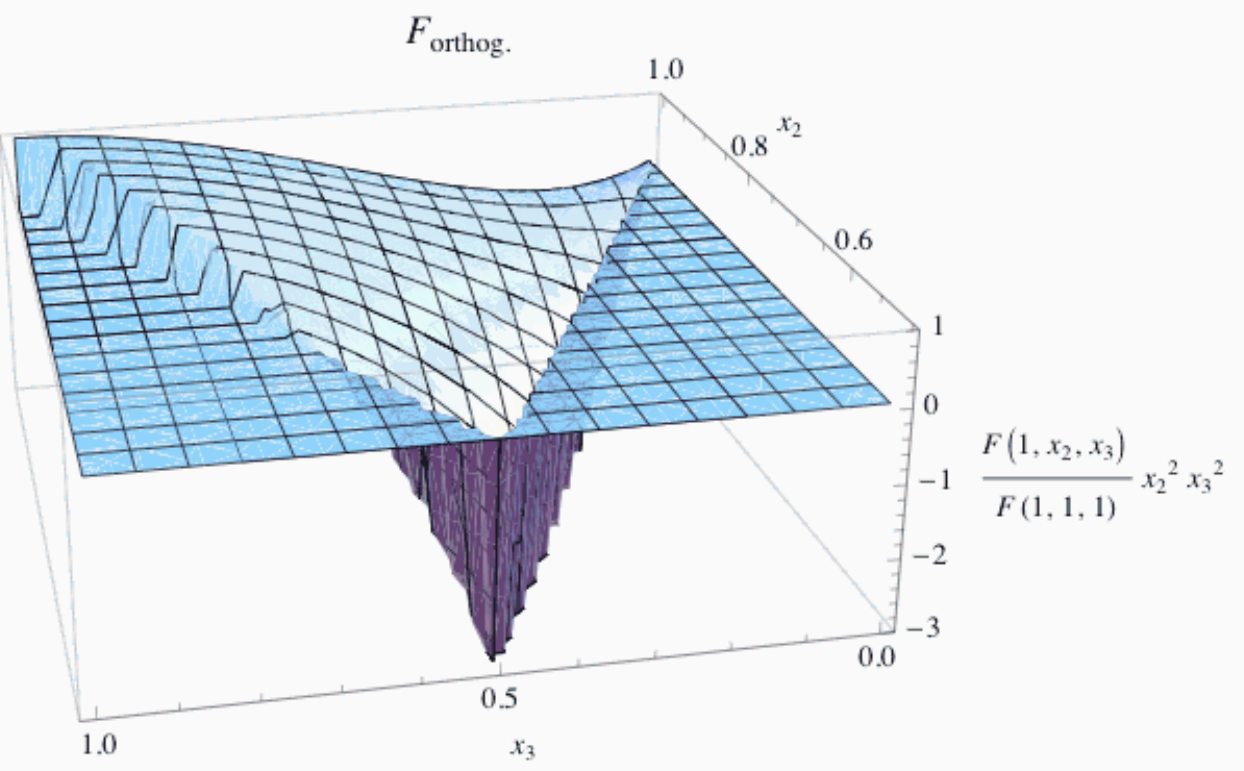}
\includegraphics[width=8.2cm]{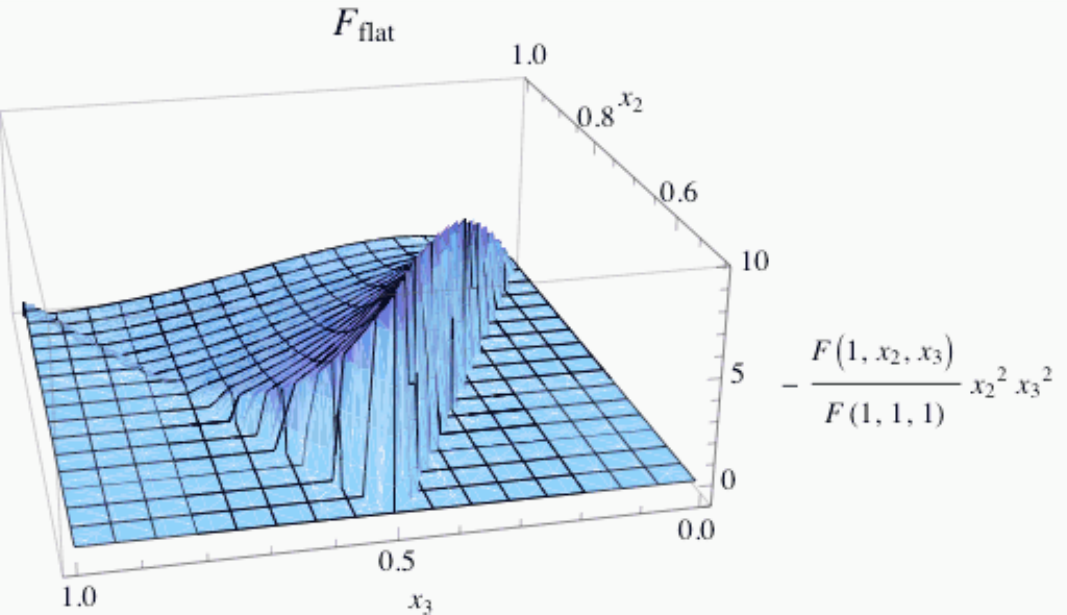}
\includegraphics[width=8.2cm]{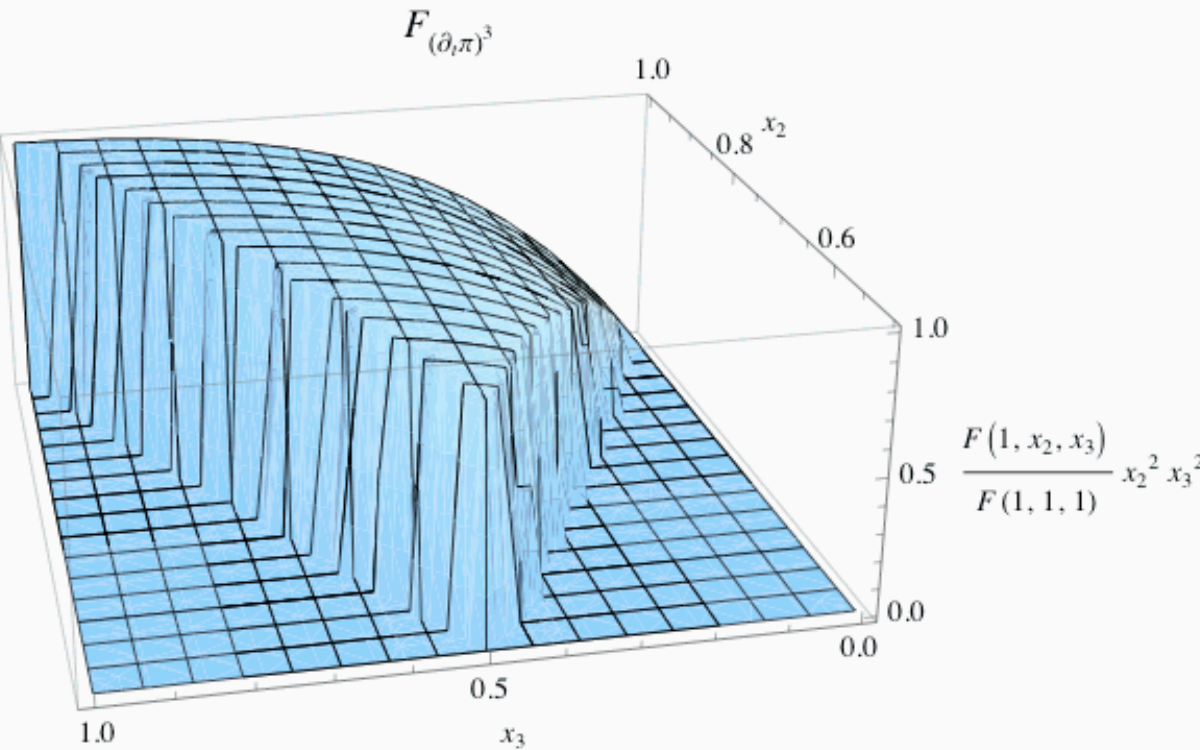}
\caption{\small The shape of single-field inflation. {\it Top Left:} $F_{\dot\pi(\d_i\pi)^2}$ (corresponding to $\tilde c_3=0$), which is very similar to the template Equilateral shape. {\it Top Right:} Orthogonal shape: $\tilde c_3=-5.4$. The cosine of this shape with the equilateral shape is approximately zero.  {\it Bottom Left:} Flat shape: $\tilde c_3=-6$. This shape is peaked on flat triangles where the two smallest $k$'s are equal to half the larger one, instead of on equilateral triangles. {\it Bottom Right:} $F_{\dot\pi^3}$, which correponds to the case $1\ll |\tilde c_3|\lesssim{\cal{O}}(10)$: the contribution on flat triangles is clearly larger than in the case of $F_{\dot\pi(\d_i\pi)^2}$.}
\label{fig:shape}
\end{figure}

\begin{figure}[htp]
\centering
\includegraphics[width=8.2cm]{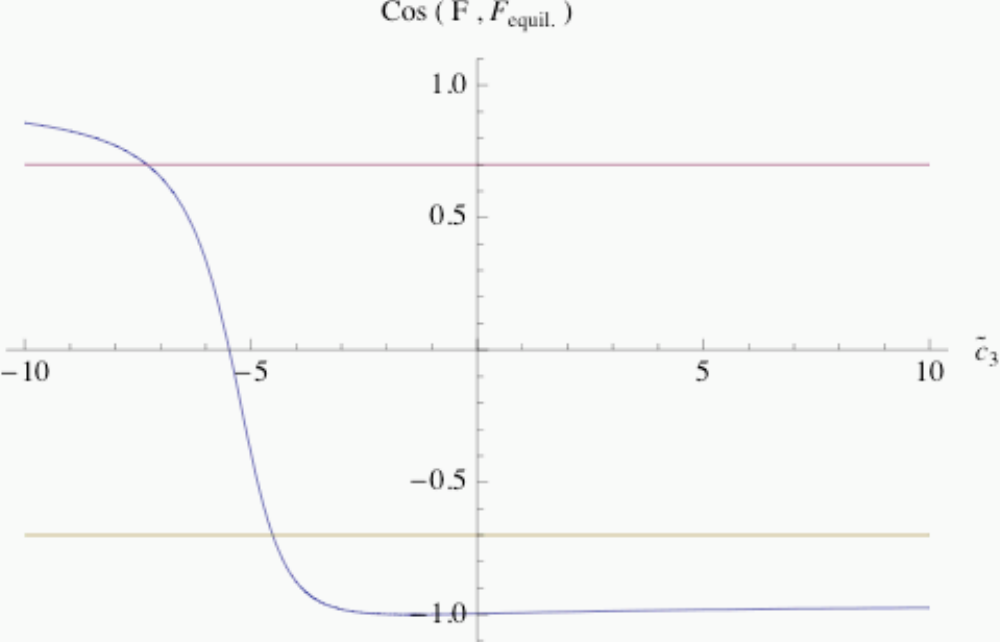}
\includegraphics[width=8.2cm]{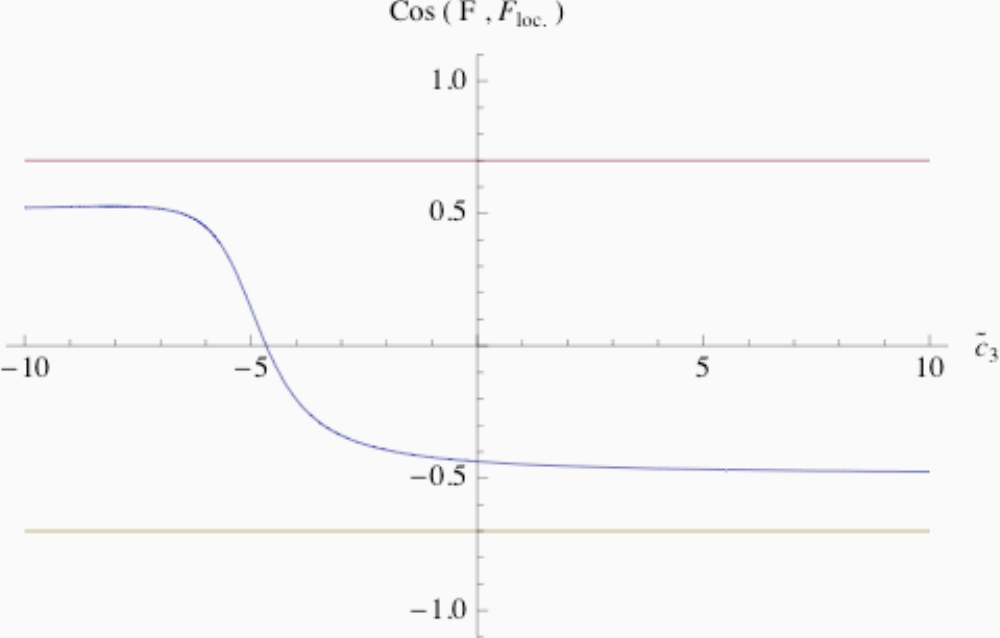}
\caption{\small {\it Left:} Cosine of single-field shape with the equilateral shape as we vary $\tilde c_3$ with $c_s\ll1$, the regime in which it is independent of $c_s$. The two horizontal lines represent when the scalar product is equal to $\pm0.7$, to give a rough measure of when the cosine becomes small. {\it Right:} Cosine with the local shape.}
\label{fig:scalar_product}
\end{figure}

\begin{figure}[htp]
\centering
\includegraphics[width=10.2cm]{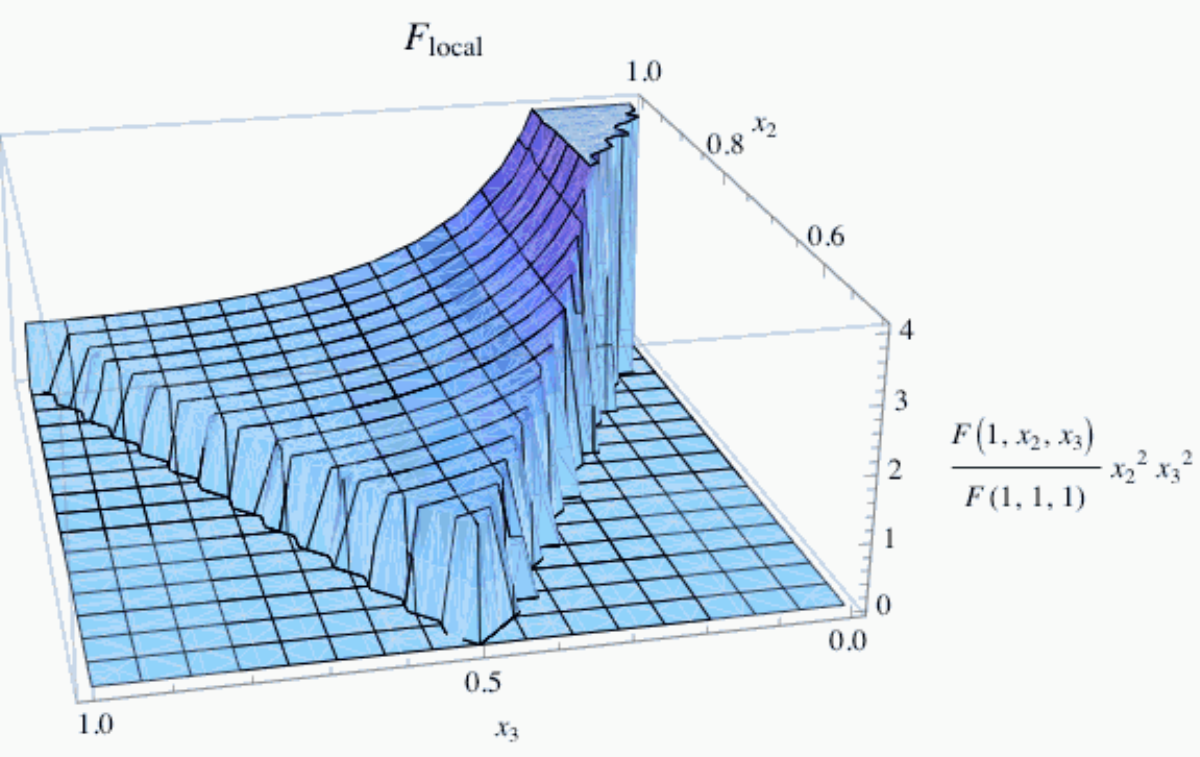}
\caption{\small The local shape.}
\label{fig:local_shape}
\end{figure}

We can use the standard definition of $f_{NL}$
\be
F(k,k,k)=f_{NL}\cdot\frac{6\Delta_\Phi^2}{k^6} \ ,
\ee
 to define
\bea\label{eq:newfnl}
&&f_{NL}^{\dot\pi(\d_i\pi)^2}=\frac{85}{324}\left(1-\frac{1}{c_s^2}\right) \ , \\ \nonumber
&&f_{NL}^{\dot\pi^3}=\frac{10}{243}\left(1-\frac{1}{c_s^2}\right)\left(\tilde c_3+\frac{3}{2}c_s^2\right)\ ,
\eea
and to write
\bea\label{eq:shapeone}
&&F_{\dot\pi(\d_i\pi)^2}(k_1,k_2,k_3)=-\frac{27}{17}\; f_{NL}^{\dot\pi(\d_i\pi)^2} \Delta_\Phi^2 \\ \nonumber
&&\quad\qquad\times\frac{\left(
24 K_3{}^6- 8 K_2{}^2 K_3{}^3 K_1- 8 K_2{}^4 K_1{}^2+22 K_3{}^3 K_1{}^3- 6 K_2{}^2 K_1{}^4+ 2 K_1{}^6\right)}{K_3{}^9K_1{}^3}\ , \\ \nonumber
&&F_{\dot\pi^3}(k_1,k_2,k_3)=162\; f_{NL}^{\dot\pi^3} \Delta_\Phi^2\cdot\frac{1}{K_3{}^3K_1{}^3}\ .
\eea 
Notice that, quite remarkably, if we wish to have a speed of sound smaller than one (a necessary condition for the existence of a Lorentz invariant UV completion~\cite{Adams:2006sv}), we need to have $f_{NL}^{\dot\pi(\d_i\pi)^2}<0$.
Even more importantly, the size and the shape of the three-point function is controlled by two parameters: the speed of sound $c_s$ and the parameter $\tilde c_3$, and the resulting non-Gaussian signal is a linear combination of two independent shapes. This means that if we fix the size of the overall non-Gaussianity there is a {\it one-parameter} family of shapes associated to the same amount of signal. In particular, for $\tilde c_3$ of ${\cal{O}}(1)$, a necessary condition for the 3-point function to be large is that the speed of sound is small \footnote{Another possibility is to be very close to the de Sitter limit. We will comment on this later.}. Given the current experimental sensitivity, we can therefore concentrate on this limit.
As we let $\tilde c_3$ vary keeping $c_s$ fixed the size and the shape of the produced non-Gaussianity change. Concentrating on the shape, for very large or very small values of $\tilde c_3$ the non-Gaussianity is dominated by one of the two operators, which both have a shape quite close to the so called equilateral kind. In that case the signal is concentrated on equilateral triangular configurations of the sort shown in the top-left panel of Fig.~\ref{fig:shape}. However the two shapes are not identical which implies that there is a region at intermediate values of $\tilde c_3$ where the shape of the resulting three-point function is completely different, being peaked on flat triangles where the size of the two lowest momenta is precisely half of the highest one. This can be seen in Fig.~\ref{fig:shape} where we show the shape of the non gaussianity for several values of the parameter $\tilde c_3$. More quantitatively, in Fig.~\ref{fig:scalar_product} we show the cosine between  the generic shape $F$ generated by single-field inflation (far from the near-de-Sitter limit) with $F_{\dot\pi(\d_i\pi)^2}$ as we let $\tilde c_3$ vary. The cosine  between two shapes was defined in \cite{Babich:2004gb}, and it represents a quantitative measure of the similarity and the correlation of the signals. Given two shapes $F_{(1)}(k_1,k_2,k_3)$ and $F_{(2)}(k_1,k_2,k_3)$, one first defines a 3-dimensional scalar product between the shapes as:
\be\label{eq:scalar_product}
F_{(1)}\cdot F_{(2)}=\sum_{k_i^{\rm physical}} F_{(1)}(k_1,k_2,k_3)F_{(2)}(k_1,k_2,k_3)/\left(P_{k_1}P_{k_2}P_{k_3}\right)\ ,
\ee
where $P(k)$ represents the power spectrum and $k_i^{\rm physical}$ means that only the $\vec k$'s that form a triangle are included. The 3D cosine between two shapes is then defined as
\be\label{eq:3D cosine}
\cos(F_{(1)},F_{(2)})=\frac{F_{(1)}\cdot F_{(2)}}{(F_{(1)}\cdot F_{(1)})^{1/2}(F_{(2)}\cdot F_{(2)})^{1/2}}\ .
\ee From Fig.~\ref{fig:scalar_product}, we see that the cosine is very small around $\tilde c_3\simeq-5$ for a region of approximately $\sim10-20\%$ of the parameter space. Here, somewhat arbitrarily, we consider  that the natural parameter range for $\tilde c_3$ is between -10 and 10, and that  the region where the cosine is small is defined as the region where this is smaller than $0.7$. We realize that this is approximately the relevant number once we plot the scalar product of the shape with the local shape produced by multifield models \cite{Lyth:2002my,Zaldarriaga:2003my} and by the new ekpyrotic universe~\cite{Creminelli:2007aq}. This shape is plotted in Fig.~\ref{fig:local_shape}. Although the shapes are clearly very different, we see that the cosine with the equilateral shape ($\tilde c_3\simeq 0$) is approximately $0.4$. In summary for roughly 10-20\% of the natural parameter space for the non-Gaussianities in single-field inflation the shape is very different from the equilateral one. 

We find that the cosine with the local shape is also very small in the same region ($\tilde c_3\simeq-5$) where the shape is different from equilateral.  As a result both the analysis of $\fnlloc$ and $\fnlequil$ that have been carried on so far, have been largely insensitive to this region of parameter space and therefore there could be an undetected large  signal in those triangular configurations. Notice also that in Fig.~\ref{fig:shape} the shape for which the scalar product in (\ref{eq:scalar_product}) is exactly orthogonal to the equilateral shape is peaked on both equilateral and on flat triangles, with opposite sign. This ensures that  in the scalar product with the equilateral shape, there is a cancellation and the result is zero. Though the region where the shape of the non-Gaussianity is very different from the equilateral one is not large, being due to a partial cancellation of the two shapes  $F_{\dot\pi(\d_i\pi)^2}$ and  $F_{\dot\pi^3}$, we consider it to be still roughly an ${\cal{O}}(1)$ fraction of the parameter space which deserves to be explored.

One general characteristic of the shape of non-Gaussianities generated by a single-field inflation is that unless a large deviation from scale invariance is detected in the power spectrum the signal is always small in the squeezed triangle limit, where the lowest of the $k$'s is much smaller than the other two. This is the consequence of a theorem that says that the signal in that limit is proportional to the tilt of the power spectrum \cite{Maldacena:2002vr,Creminelli:2004yq,Cheung:2007sv}.    We stress that the theorem of \cite{Maldacena:2002vr,Creminelli:2004yq,Cheung:2007sv} does not apply to inflationary models where fields other than the inflaton play a relevant role, either by producing entropy perturbations \cite{Lyth:2002my,Zaldarriaga:2003my,Langlois:2008wt}, or by slowing down the field with interactions \cite{Green:2009ds,Moss:2007cv}. These models are allowed to produce large non-Gaussianties whose shape  is peaked on squeezed configurations (like the one in Fig.~\ref{fig:local_shape}), but they can also produce a signal peaked on equilateral or even flat configurations, as in \cite{Green:2009ds}.
  
In \cite{Chen:2006nt,Holman:2007na}, it was argued that modifications of the initial state of the inflaton field fluctuations can induce a departure from Gaussianity, and the shape of the induced signal, whose size strongly depends on the cutoff, was computed. Though we find the theoretical motivation for these models quite unclear, the resulting shape of the signal is very similar to the one produced by single field inflation with a particular value of $\tilde c_3$. Therefore, as we will explain in the next section, our analysis  will automatically cover the  signal expected in this case as well.

Summarizing, the general analysis using the effective field theory of inflation shows us that at least when the system is away from de Sitter the non-Gaussian signal generated by single-field inflation when an approximate shift symmetry protects the Goldstone boson is a linear combination of two independent shapes. In general the resulting signal can be very different from the one associated with the equilateral shape. It therefore requires a more complete analysis that we will soon undertake. As we are now going to see, this is the case also when the system is close to de Sitter space.

\subsection{Near-de-Sitter limit\label{sec:EFTds}}

When the inequality (\ref{eq:ds_limit}) is satisfied, then the kinetic terms in the last line of (\ref{eq:Lagrangian}) become the dominant ones at horizon crossing.  In this case, it is a good and useful approximation to take the exact de Sitter limit, sending $\dot H$ and $c_s$ in (\ref{eq:Lagrangian}) to zero in such a way that the quantity $\mpl^2 \dot H(1-c_s^2)/c_s^2$ approaches a constant value
\be
-\frac{\dot H\mpl^2}{c_s^2}(1-c_s^2)\rightarrow 2M^4\ .
\ee
Notice that the new parameter $M$ is the same as defined in \cite{Cheung:2007sv}. In this regime, the standard spatial kinetic term of the first line of (\ref{eq:Lagrangian}) goes to zero, while the ones on the third line become important. The two kinetic operators possible in this regime are important in different regimes depending on the size of the ratio $4 (d_2+d_3)^{1/2}/|d_1|$.  We will concentrate on the cases where only one of the two operators dominate at horizon crossing, neglecting the quite tuned case where both of the operators contribute at the same time, since, as it will become clear later, in this case we do not expect any new qualitative feature in the shape of the resulting non-Gaussianity.

In the limit when $|d_1|\lesssim 4(d_2+d_3)^{1/2}$, then it is the higher-derivative spatial-kinetic term that dominates at horizon crossing. This is the regime of Ghost inflation  \cite{Cheung:2007st,Cheung:2007sv,ArkaniHamed:2003uz,Senatore:2004rj}, that here we are able to describe as a continuous deformation from more standard models. In this case, the dispersion relation of the $\pi$ mode becomes extremely non-relativistic:
\be\label{eq:non-linear}
\omega\propto \frac{k^2}{M}\ .
\ee
Given the non-linear dispersion relation, the way an operator scales with energy does not coincide with its mass dimension as in the Lorentz invariant case. A rescaling of the energy by a factor $s$, $E \rightarrow s\,E$, (equivalent to a time
rescaling $t \rightarrow s^{-1}t$), must go together with a momentum transformation $k \rightarrow s^{1/2}k$ ($x \rightarrow s^{-1/2}x$ on the spatial coordinates) and a $\pi$ transformation $\pi\rightarrow s^{1/4}\pi$. Since in making predictions for inflation, we are interested in energy scales of order $H$, there are two interaction operators which are the most relevant (or, in technical language, the least irrelevant): $\dot\pi (\d_i\pi)^2$ and $(\d_j^2\pi)(\d_i\pi)^2$, which have scaling dimension equal to $1/4$. For example, the operator $\dot\pi^3$ we considered before has scaling dimension $5/4$ and it therefore gives rise to an effect much smaller than the one induced by $\dot\pi (\d_i\pi)^2$ and $(\d_j^2\pi)(\d_i\pi)^2$. The form of the non-Gaussianity these two operators induce is rather complicated, as the non-linear dispersion relation of (\ref{eq:non-linear}) makes the wavefunction of $\pi$, and therefore the resulting shape of the induced non-Guassianity,  very complicated and not-writable in a closed form. The calculation can be easily carried on following the steps presented in \cite{Cheung:2007st,ArkaniHamed:2003uz,Senatore:2004rj}, where the non-Gaussianity induced by the operator  $\dot\pi (\d_i\pi)^2$ is computed. The wavefunction for $\pi$ is given by~\cite{Cheung:2007st,ArkaniHamed:2003uz}:
\be\label{eg:ghost_wavefunction}
\pi^{cl}_k(\tau)=-\sqrt{\frac{\pi}{8}}\frac{H}{2 M^2}(- \tau)^{3/2} H^{(1)}_{3/4}\left(\frac{k^2H d_3^{1/2}}{4M}\tau^2\right)\ .
\ee
where $\tau$ is the conformal time, and 
$H^{(1)}_\nu$ is the Hankel function of the first kind.
Here we have expanded the operator $\pi$ in creation and annihilation operators in the usual way $\pi_k=\pi_k^{cl}\hat a_k+\pi^{cl}_{-k}{}^*\, \hat a^\dag$.
The contribution to the 3-point function from the operator $\dot\pi(\d_i\pi)^2$ is given by the usual expression
\bea\label{eq:ghost1}
\langle\Phi_{\vec k_1}\Phi_{\vec k_2}\Phi_{\vec k_3}\rangle_{\dot\pi(\d_i\pi)^2}\equiv &&(2\pi)^3\delta^{(3)}\left(\sum_{i}\vec k_i\right)  \bar F_{\dot\pi(\d_i\pi)^2}=\\ \nonumber 
&&-i\; (2\pi)^3\delta^{(3)}\left(\sum_{i}\vec k_i\right) \left(\frac{3}{5}\right)^3 H^3{{C}_{\dot\pi(\d_i\pi)^2}} \pi^{cl}_{k_1}(0) \pi^{cl}_{k_2}(0) \pi^{cl}_{k_3}(0)\\ \nonumber
&&\cdot\int^0_{-\infty}\frac{d\tau}{H\tau}\frac{d}{d\tau}\pi_{k_1}^{cl}{}^*(\tau)\pi_{k_2}^{cl}{}^*(\tau)\pi_{k_3}^{cl}{}^*(\tau)(\vec k_2\cdot \vec k_3)+{\rm permutations} + {\rm c.c.}\ ,
\eea
where the sum above includes all symmetric permutations of the three momenta and the contour of integration should be rotated into the complex plane to ensure convergence as $\tau\rightarrow-\infty$.
${{C}_{\dot\pi(\d_i\pi)^2}}$ is the coefficient in the Lagrangian (\ref{eq:Lagrangian}) of the operator $\dot\pi(\d_i\pi)^2/a^2$, which is equal in this case to
\be
C_{\dot\pi(\d_i\pi)^2}=-2M^4\ .
\ee

Analogously for the operator $(\d_i^2\pi)(\d_j\pi)^2$ we have:
\bea\label{eq:ghost2}
\langle\Phi_{\vec k_1}\Phi_{\vec k_2}\Phi_{\vec k_3}\rangle_{(\d_i^2\pi)(\d_j\pi)^2}\equiv &&(2\pi)^3\delta^{(3)}\left(\sum_{i}\vec k_i\right)  \bar F_{(\d_i^2\pi)(\d_j\pi)^2}\\ \nonumber =&&-i\; (2\pi)^3\delta^{(3)}\left(\sum_{i}\vec k_i\right) \left(\frac{3}{5}\right)^3 H^3{{C}_{(\d_i^2\pi)(\d_j\pi)^2}} \pi^{cl}_{k_1}(0) \pi^{cl}_{k_2}(0) \pi^{cl}_{k_3}(0)\\ \nonumber
&&\cdot\int^0_{-\infty}d\tau\, k_1^2\pi_{k_1}^{cl}{}^*(\tau)\pi_{k_2}^{cl}{}^*(\tau)\pi_{k_3}^{cl}{}^*(\tau)(\vec k_2\cdot \vec k_3)+{\rm permutations} + {\rm c.c.}\ ,
\eea
where $C_{(\d_i^2\pi)(\d_j\pi)^2}$ is in this case given by
\be
C_{(\d_i^2\pi)(\d_j\pi)^2}=-\frac{d_1 M^3}{4}\ .
\ee
The total 3-point function is given in this case by
\be
\langle\Phi_{\vec k_1}\Phi_{\vec k_2}\Phi_{\vec k_3}\rangle=(2\pi)^3\delta^3(\sum_i\vec k_i)\left(\bar F_{\dot\pi(\d_i\pi)^2}+\bar F_{(\d_j^2\pi)(\d_i\pi)^2}\right)
\ee

Unfortunately, due to the complicated form of the wavefunction in (\ref{eg:ghost_wavefunction}), there is no closed form expression for the above integrals, which need to be integrated numerically.
What is interesting to us for what concerns the data analysis, is that the resulting two independent shapes $ \bar F_{\dot\pi(\d_i\pi)^2}$ and $\bar F_{(\d_i^2\pi)(\d_j\pi)^2}$ are peaked on equilateral configuration, but still  different. This means that the two-parameter space of non-Gaussianities they generate leads to a signal that, in some region, is very different from the equilateral kind, peaked on flat-triangular configurations.
The values of the $f_{NL}$ parameters they induce, defined in the usual way, are given by:
 \bea\label{eq:newfnlghost}
&&f_{NL}^{\dot\pi(\d_i\pi)^2}\simeq0.2547 \frac{1}{(d_2+d_3)^{1/2}}\cdot\frac{M}{H}\simeq 138.1\cdot \frac{1}{(d_2+d_3)^{4/5}} \ , \\ \nonumber
&&f_{NL}^{(\d_j^2\pi)(\d_i\pi)^2}\simeq0.1327 \frac{d_1 }{ (d_2+d_3) }\cdot\frac{M}{H}\simeq 71.97 \cdot \frac{d_1}{(d_2+d_3)^{13/10}} \ .
\eea
where  we have used the normalization of the power spectrum
\be
\Delta_\Phi=\frac{9\,\pi\,\sqrt{2}}{25\,\Gamma(1/4)^2}\cdot\frac{1}{(d_2+d_3)^{3/4}}\left(\frac{H}{M}\right)^{5/4}. 
\ee
These expressions are valid when $|d_1|\lesssim4 (d_2+d_3)^{1/2}$ and the inequality in~(\ref{eq:ds_limit}) is satisfied. Notice that they imply that the speed of sound of the fluctuations, meant as the coefficient of the linear term in $k$ in the dispertion relation at energies higher than $H$,  should be bounded by 
\be
|c_s|\lesssim 10^{-2}(d_2+d_3)^{2/5}\ .
\ee
where the absolute value is used since for $d_1<0$, $c_s^2$ is negative.

When instead $d_1\gtrsim4 (d_2+d_3)^{1/2}$, then the spatial kinetic term at horizon crossing is a standard two derivative one, giving a speed of sound for the fluctuations equal to \cite{Cheung:2007st,Cheung:2007sv,Creminelli:2006xe}:
\be
c_s^2=\frac{d_1 H}{8 M}\ .
\ee
We will deal here with the case of positive $d_1$, and come back to the case of $d_1<0$ in the next subsection.  Because of the extremely low speed of sound, the higher-derivative tri-linear operator of the last line of (\ref{eq:Lagrangian}) becomes as important as the ones we considered in the former section. The shape that is generated by this operator is this time writable in closed form, and it has the following form
\bea\label{eq:Fhigherderiv}
&&F_{(\d_j^2\pi)(\d_i\pi)^2}(k_1,k_2,k_3)=\frac{5}{3}\frac{1}{c_s^2}\cdot\Delta_\Phi^2
 \\ \nonumber
&&\quad\qquad\times\frac{\left(
24\, K_3{}^6- 4\,  K_2{}^2 K_3{}^3 K_1- 4\,  K_2{}^4 K_1{}^2+11\,  K_3{}^3 K_1{}^3- 3\,  K_2{}^2 K_1{}^4+  K_1{}^6\right)}{K_3{}^9K_1{}^3}\ . \\ \nonumber 
\eea
Notice that, quite surprisingly, this shape is a linear combination of the shapes $F_{\dot\pi(\d_i\pi)^2}$ and $F_{\dot\pi^3}$:
\be
F_{(\d_j^2\pi)(\d_i\pi)^2}=\frac{5}{3}\frac{1}{c_s^2}\cdot \Delta_\Phi^2\left.\left(-\frac{17}{54}F_{\dot\pi(\d_i\pi)^2}+\frac{6}{81} F_{\dot\pi^3}\right)\right|_{f_{NL}\cdot\Delta_\Phi^2=1}
\ee
The non-gaussianity induced by the other two operators $\dot\pi(\d_i\pi)^2$ and $\dot\pi^3$ is as in eq.~(\ref{eq:shapezero}), which leads us to conclude that in this case the 3-point function is given by
\be
\langle\Phi_{\vec k_1}\Phi_{\vec k_2}\Phi_{\vec k_3}\rangle=(2\pi)^3\delta^3(\sum_i\vec k_i)\left(F_{\dot\pi(\d_i\pi)^2}+ F_{\dot\pi^3}+F_{(\d_j^2\pi)(\d_i\pi)^2}\right)
\ee
with the values of the $f_{NL}$'s, defined in the usual way, given by:
\bea\label{eq:newfnlcs}
&&f_{NL}^{\dot\pi(\d_i\pi)^2}=-\frac{85}{324}\cdot\frac{1}{c_s^2}\simeq -2.662 \cdot10^3 \cdot\frac{1}{d_1^{8/5}} \ , \\ \nonumber
&&f_{NL}^{\dot\pi^3}=-\frac{10}{243}\left(\frac{\tilde c_3}{(d_2+d_3)^4}+\frac{3}{2}\right)\simeq -4.115\cdot10^{-2}\cdot\frac{\tilde c_3}{(d_2+d_3)^4}
\ , \\ \nonumber
&&f_{NL}^{(\d_j^2\pi)(\d_i\pi)^2}=-\frac{65}{162}\cdot \frac{1}{c_s^2}\simeq- 4.072 \cdot 10^3 \cdot\frac{1}{d_1^{8/5}}\ .
\eea
where  we have used the normalization of the power spectrum, which in this limit has the peculiar form of
\be \label{eq:near_desitter_cs}
\Delta_\Phi=\frac{4608}{25}\frac{c_s^5}{d_1^4}\ .\ee
A notable difference with respect to (\ref{eq:shapezero}) is the coefficient of the operator $\dot \pi^3$.  This is due to the fact that at energies higher than the crossing energy $E_{\rm cr}=d_1 H/(d_2+d_3)^{1/2}$, the spatial kinetic term is dominated by the $k^4$ term. This means that the scaling dimensions of the operators $\dot\pi^3$ and $\dot\pi(\d_i\pi)^2$ become different, with the operator $\dot\pi^3$ having scaling dimension equal to $5/4$ while  $\dot\pi(\d_i\pi)^2$ has scaling dimension equal to $1/4$. Since the cutoff of the theory in this case is given by $\Lambda\sim M (d_2+d_3)^{7/2}$, it is quite straightforward to see that the operator $\dot\pi^3$ becomes strongly coupled at the same scale as $\dot\pi(\d_i\pi)^2$ if we substitute
\be\label{eq:c3tilde_scaling}
\frac{\tilde c_3}{c_s^2}\rightarrow \frac{\tilde c_3}{c_s^2}\left(\frac{E_{\rm cr}}{\Lambda}\right)\sim\frac{\tilde c_3}{(d_2+d_3)^4}\ ,
\ee
with $\tilde c_3$ of order one. Unless $d_2+d_3$ is small, the non-Gaussianity induced by this operator is rather negligible.

We have three operators generating non-Gaussianities, but only two independent coefficients (which means we have only a bi-dimensional space of non-Gaussianity). This can be clearly realized by writing the Lagrangian (\ref{eq:Lagrangian}) in terms of $c_s$: $d_1$ disappears, and one is left with an expression of the form (neglecting numerical coefficients)
\be
M^4(\dot\pi^2-c_s^2(\d_i\pi)^2)+M^4\dot\pi(\d_i\pi^2)+M^4\frac{\tilde c_3}{(d_2+d_3)^4}\dot\pi^3+M^4 \frac{c_s^2}{H}(\d_j^2\pi)(\d_i\pi^2)\ .
\ee
Since non-Gaussianities are generated when the modes cross the horizon, a time derivative contributes as a factor of $H$, while a spatial derivative as $H/c_s$. Taking this into account, it is immediate to see that the operators $\dot\pi(\d_i\pi)^2$ and $(\d_j^2\pi)(\d_i\pi)^2$ give rise to an $f_{NL}$ which is parametrically the same ${\cal{O}}(1/c_s^2)$.
Notice again that the above expressions for $f_{NL}$'s are valid only when $d_1\gtrsim4( d_2+d_3)^{1/2}$ and the inequality~(\ref{eq:ds_limit}) is satisfied.

Summarizing, the induced non-Gaussianity when we are close to the de Sitter limit can be large and detectable. The effective field theory shows that, depending on which operator dominates at horizon crossing there are two independent shapes of non-Gaussianities. As we concluded in the case where the system is less close  to de Sitter, a more complete analysis is required.

\subsection{Negative $c_s^2$\label{sec:negatice_cssq}}

There is one last regime which needs to be explored: the case where the squared speed of sound of the fluctuations $c_s^2$ is negative. The speed of sound is given by:
\be\label{eq:speed_of_sound_negaritve}
c_s^2=\frac{-\dot H \mpl^2+d_1 H M^3/4}{2 M^4-\dot H \mpl^2-3d_1 H M^3/2}\ ,
\ee
where here we have not assumed that $\dot H$ is negligible. Imposing the time-kinetic term to be positive forces the inequality $2M^4>\dot H\mpl^2+3d_1 H M^3/2$.
We see that $c_s^2$ can be negative when $\dot H$ is positive or $d_1$ is negative. In this case, the dispersion relation, deep inside the horizon, is given by
\be
\omega^2=-|c_s^2| k^2+\frac{(d_2+d_3)}{4 M^2}k^4\ .
\ee
where in the last term we have neglected a generally small correction proportional to $\dot H \mpl^2$ and to $d_1 H M^3$.
 When this happens, and the term in $c_s^2 k^2$ begins to dominate over the term proportional to $k^4$ at horizon crossing, the system is unstable, and the modes begin to grow exponentially. This does not mean the inflationary model is inconsistent. In fact, if the term in $k^4$ is large enough to dominate  deep inside the horizon before the cutoff scale, the modes are stable in the ultraviolet, and then, as they redshift down to the Hubble scale, they become unstable and begin to grow exponentially. However, they do so  only in a window of energies from when the term in $c_s^2 k^2$ dominates down to the Hubble scale, when the modes freeze out and stop to grow. This means that such inflationary models are consistent. One interesting feature of these models is that they allow, if $\dot H$ is positive, to have a consistent violation of the null energy condition, and an inflationary model with a potentially detectable blue tilt of gravity waves \cite{Senatore:2004rj,Creminelli:2006xe}~\footnote{Notice that $\dot H$ can be positive but small enough so that there is no exponential growth of the modes before horizon crossing \cite{Creminelli:2006xe}. This model would still imply a potentially measurable blue tilt of gravity waves. For what concerns the power spectrum and the non-Guassianities, this model is included in what studied in sec.~\ref{sec:EFTds}.}.

Notice however that, as we will see, the fact that the modes grow exponentially for a window of time makes the induced non-Gaussianities rather large, and a consistent fraction of the parameter space of these models, as we will verify in the next section, is ruled out. This is why the operator in $k^4$ has to dominate soon enough at high energies, which explains why we still need to be close to the de Sitter regime, with the inequality (\ref{eq:ds_limit}) not  violated by too much. 

We will concentrate on two cases separately: when the $|d_1|\gtrsim \frac{4\dot H \mpl^2}{M^3 H}$, with $d_1<0$, then the speed of sound is approximately given by
\be
c_s^2\simeq\frac{d_1  H }{8 M}<0\ ,
\ee
and then in the case when  $|d_1|\lesssim \frac{4\dot H \mpl^2}{M^3 H}$, with $\dot H>0$, in which case
\be\label{eq:cs_negative_2}
c_s^2\simeq-\frac{\dot H \mpl^2}{2 M^4}<0\ .
\ee
Here we have assumed $c_s\ll 1$ and $d_1 H\ll M$ for simplicity.
In the first case, as we saw in the former subsection, the condition for the term in $c_s^2 k^2$ to dominate over the one in $k^4$ implies $d_1\lesssim -4\,(d_2+d_3)^{1/2}$, where, as usual, the ${\cal{O}}(1)$ coefficients are not under control. The calculation for the power spectrum and the non-Gaussianities follows very closely the one of  \cite{Senatore:2004rj}, and we do not give the details here. Deep inside the horizon, the dispersion relation is dominated by the stable $k^4$ term, which defines a stable vacuum, and the wave function agrees with the one of Ghost inflation (\ref{eg:ghost_wavefunction}). As far as the non-Gaussianities are concerned, since the modes at horizon crossing are dominated by a linear dispersion relation, the shape of the non-Gaussianities is exactly equal to the one with a positive 
$c_s^2$ given in eq.~(\ref{eq:shapeone}) and~(\ref{eq:Fhigherderiv}), with different values for the $f_{NL}$'s. These are given by:
\bea\label{eq:fnlcs_negative_1}
f_{NL}^{\dot\pi(\d_i\pi)^2}&=&\frac{85}{324}\frac{1}{|c_s|^2}e^{-\frac{d_1}{2 (d_2+d_3)^{1/2}}}=2.528 \cdot 10^{-11}\frac{ d_1^4}{|c_s|^7}\ ,\\ \nonumber
f_{NL}^{(\d_j^2\pi)(\d_i\pi)^2}&=&\frac{65}{162}\frac{1}{|c_s|^2}e^{-\frac{d_1}{2 (d_2+d_3)^{1/2}}}=3.867\cdot 10^{-11}\frac{ d_1^4}{|c_s|^7}\ ,
\eea
where in the second passage we have used that the normalization of the power spectrum is given by:
\be\label{eq:deltacs_negative_1}
\Delta_\Phi=\frac{4608}{25} \frac{|c_s|^5}{d_1^4} e^{-\frac{d_1}{2 (d_2+d_3)^{1/2}}}\ .
\ee
Notice, as expected, the exponential dependence on the ratio $-d_1/{(d_2+d_3)^{1/2}}$, with $d_1<0$, that controls the amount of exponential grow of the modes before horizon crossing. Here we have assumed that the contribution from the operator $\dot\pi^3$ is irrelevant. This is justified by the fact that, as we have seen in eq.~(\ref{eq:c3tilde_scaling}), its expected importance decreases as we decrease the interval in energies between when the  $k^4$ term dominates and when the modes crosses the horizon. This is the interval during which the term in $c_s^2 k^2$ dominates. Because of the exponential dependence of $f_{NL}$ on this same interval of energies, we expect (and we will later verify) that this interval has to be rather small, and that therefore the operator $\dot\pi^3$ gives a negligible contribution. 

The situation in the case where $|d_1|\lesssim \frac{4\dot H \mpl^2}{M^3 H}$, with $\dot H>0$, is very similar. The condition for the $c_s^2 k^2$ term to dominate at horizon crossing becomes $(d_2+d_3)\lesssim 8 M^6/(H^2\dot H \mpl^2) $.
For the non-Gaussianities, we obtain:
\bea\label{eq:fnlcs_negative_2}\nonumber
f_{NL}^{\dot\pi(\d_i\pi)^2}&=&\frac{85}{324}\frac{1}{|c_s|^2}{\rm Exp}\left[{2^{7/4}\left(\frac{\dot H \mpl^2}{H^4}\right)^{1/4}\frac{|c_s|^{3/2}}{d_3^{1/2}}}\right]=5.179 \cdot 10^{-8}\frac{1}{|c_s|} \left(\frac{\dot H \mpl^2}{H^4}\right)\, ,\\ \nonumber
f_{NL}^{(\d_j^2\pi)(\d_i\pi)^2}&=&\frac{65}{162}\frac{1}{|c_s|^2}\frac{-d_1}{\left(\frac{4\dot H \mpl^2}{M^3 H}\right)}{\rm Exp}\left[{2^{7/4}\left(\frac{\dot H \mpl^2}{H^4}\right)^{1/4}\frac{|c_s|^{3/2}}{d_3^{1/2}}}\right]=\\ \nn
&=&-1.177\cdot 10^{-8}\frac{d_1}{|c_s|^{5/2}}\left(\frac{\dot H \mpl^2}{H^4}\right)^{3/4}=\\ 
&=&-1.177\cdot 10^{-8}\frac{1}{|c_s|}\left(\frac{\dot H \mpl^2}{H^4}\right) \cdot \frac{d_1}{|c_s|^{3/2}}\left(\frac{\dot H \mpl^2}{H^4}\right)^{-1/4}\ ,
\eea
where similarly in the second passages we have used that the normalization of the power spectrum is given by:
\be
\Delta_\Phi=\frac{9}{100}  \frac{H^4}{\dot H \mpl^2}\frac{1}{|c_s|}{\rm Exp}\left[{2^{7/4}\left(\frac{\dot H \mpl^2}{H^4}\right)^{1/4}\frac{|c_s|^{3/2}}{d_3^{1/2}}}\right]\ ,
\ee
and we have similarly neglected the contribution from the operator $\dot\pi^3$.

Summarizing, we see that there are consistent inflationary models with a negative squared speed of sound at horizon crossing. They induce exponentially large non-Gaussianities with, in general, two independent shapes. As for the other regimes that we discussed in the two former sections, a more complete analysis is required.

 \section{Templates for single-field inflation non-Gaussianties\label{sec:strategy}}
 
 In the former section we have studied the non-Gaussianities that can be generated by single field inflation, under the assumption that an approximate shift symmetry protects the Goldstone boson.  We have seen that the interesting regimes can be differentiated according to wether the system is close or not to de Sitter, following the inequality in~(\ref{eq:ds_limit}). We will show however that the same data analysis technique is sufficient for all these cases.
 
 Let us start from the case where the system is not close to de Sitter.
 In this case we have seen that the non-Gaussianities are given by a linear combination of the signal induced by two operators: $\dot\pi^3$ and $\dot\pi(\d_i\pi)^2$. Each of those two operators gives rise to shapes for the non-Gaussianities that are peaked on equilateral configurations  but are still quite different. Since non-Gaussianities from a generic  single-field model of inflation which is not very close to de Sitter will be the result of a combination of the effect from these two operators, the resulting shape of the signal can be very different, peaked on flat configurations or even changing sign as we go from equilateral to flat triangular configurations.
 We therefore conclude that it is necessary to do the analysis of the non-Gaussianity jointly for both of the shapes $F_{\dot\pi(\d_i\pi)^2}$ and $F_{\dot\pi^3}$. The numerical analysis becomes much simpler if instead of doing the analysis for  precisely these two shapes, we use templates which are very similar but which, being factorizable as a product
of functions of $k_1$, $k_2$ and $k_3$ or as a sum of a small number of terms with this property, are computationally efficient \cite{Wang:1999vf}.
 For the equilateral shape, this is given by \cite{Creminelli:2005hu}:
\be
F_{\rm equil.}(k_1,k_2,k_3)=f_{NL}^{\rm equil.}\cdot6\Delta_{\Phi}^2\cdot\left(-\frac{1}{k_1^3k_2^3}-\frac{1}{k_1^3k_3^3}-\frac{1}{k_2^3k_3^3}-\frac{2}{k_1^2k_2^2k_3^2}+\frac{1}{k_1 k_2^2k_3^3}+(5\ perm.)\right)\ .
\ee
where the permutations act only on the last term in parenthesis. 
The second independent template we choose is similar to the orthogonal shape (see upper-right panel of Fig.~(\ref{fig:shape})), and it is given by:
\be\label{eq:F_orthog}
F_{\rm orthog.}(k_1,k_2,k_3)=f_{NL}^{\rm orthog.}\cdot6\Delta_{\Phi}^2\cdot\left(-\frac{3}{k_1^3k_2^3}-\frac{3}{k_1^3k_3^3}-\frac{3}{k_2^3k_3^3}-\frac{8}{k_1^2k_2^2k_3^2}+\frac{3}{k_1 k_2^2k_3^3}+(5\ perm.)\right)\ ,
\ee
where the permutations act only on the term immediately to the left. 
In practice, $F_{\rm equil.}$ is a good template for the equilateral shape with $\tilde c_3\simeq 0$, while  $F_{\rm orthog.}$  is a good template for the orthogonal shape, $\tilde c_3\simeq -5.4$. As we will later see, these two templates will be able to reproduce with good accuracy the bidimensional space spanned by the exact single-field shape as we vary the contribution of $F_{\dot\pi(\d_i\pi)^2}$ and $F_{\dot\pi^3}$ by varying $c_s$ and $\tilde c_3$. 

\begin{figure}[htp]
\centering
\includegraphics[width=11.8cm]{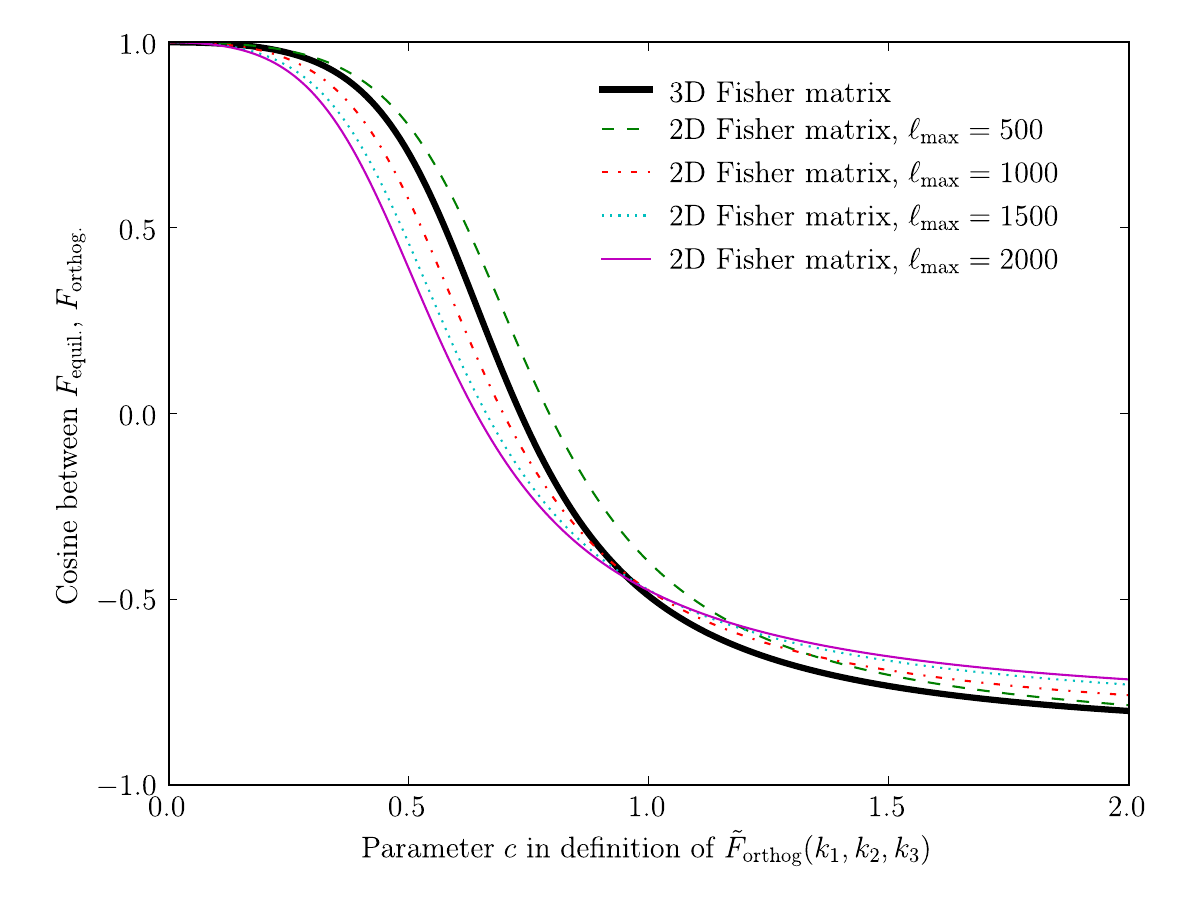}
\caption{\small Cosine between $F_{\rm equil.}$ and $\tilde F_{\rm orthog.}$ as we let $c$ vary and as we use the 3D and 2D cosine, or as we vary the $l_{\rm max}$ in the definition of the 2D cosine. As expectable, we see that how much the two shapes are similar depends on the survey. Our choice of the template $F_{\rm orthog.}$ corresponds to $c=2/3$, which is quite close to a 2D orthogonal template with repect to $F_{\rm equil.}$ at Planck resolution.}
\label{fig:cosine_c}
\end{figure}

Notice that the orthogonality of the shapes is in reality experiment dependent: we can  consider not only the three dimensional cosine but also in the case of a CMB experiment, the 2D cosine between the bispectra $B_{l_1 l_2 l_3}$ as \cite{Babich:2004gb}:
\be
\cos(B_{(1)},B_{(2)})=\frac{B_{(1)}\cdot B_{(2)}}{(B_{(1)}\cdot B_{(1)})^{1/2}(B_{(2)}\cdot B_{(2)})^{1/2}}\ ,  \label{eq:2d_cosine}
\ee
where we have defined the  2D scalar product:
\be\label{eq:2D_scalar_product}
B_{(1)}\cdot B_{(2)}=\sum_{l_1\leq l_2\leq  l_3}^{l_{\rm max}}\frac{B_{(1)}{}_{l_1 l_2 l_3}B_{(2)}{}_{l_1 l_2 l_3}}{f_{l_1l_2l_3} C_{l_1}C_{l_2}C_{l_3}}\ ,
\ee
with $f_{l_1,l_2,l_3}$ being a combinatorial factor equal to 1 if the three $l$'s are different, to 2 if two of them are equal and to 6 if all of them are equal, and with $l_{\rm max}$ being the maximum $l$ of the CMB survey. Here the bispectrum is defined in such a way as
\be
\langle a_{l_1m_1}a_{l_2m_2}a_{l_3m_3}\rangle= \left(
\begin{array}{ccc}
l_1 & l_2 & l_3 \\
m_1 & m_2 & m_3
\end{array}
\right) B_{l_1 l_2 l_3}\ ,
\ee
while
\be
\langle a_{l_1m_1} a_{l_2m_2}^*\rangle=C_{l_1}\delta_{l_1l_2}\delta_{m_1 m_2}\ ,
\ee
where the $a_{lm}$'s are the CMB multipoles. The 2D cosine is expected to give a quantitative measure of the correlation of two shapes after the transfer functions and the sky projection are applied.
If we define a more general family of orthogonal shapes
\bea\label{eq:generic_Forthog}
\tilde F_{\rm orthog.}(k_1,k_2,k_3)=f_{NL}^{\rm orthog.}\cdot\frac{1}{1-c}\Delta_{\Phi}^2\cdot \left( \left.F_{\rm equil.}(k_1,k_2,k_3)\right|_{f_{NL}^{\rm equil.}\cdot\Delta_\Phi=1}-6\frac{c}{k_1^2k_2^2k_3^2}\right)\ ,
\eea
we can consider not only the 3D cosine of eq.~(\ref{eq:3D cosine}), but also the $l_{\rm max}$-dependent 2D cosine, as we let the constant $c$ in eq.~(\ref{eq:generic_Forthog}) vary. The value of the various cosines between $\tilde F_{\rm orthog.}$ and $\tilde F_{\rm equil.}$ as we vary $c$ is shown in Fig.~\ref{fig:cosine_c}, where one can see that the value changes as we pass from 3D to 2D, and as we let $l_{\rm max}$ vary. We choose $c=2/3$ in eq.~(\ref{eq:generic_Forthog}) because this represents a good approximation to a 2D orthogonal shape for an experiment like Planck. We stress that, if one takes into account of the correlation between $\tilde F_{\rm orthog.}$ and $F_{\rm equil.}$, as we will do, all of the different choices of $c$ are equivalent, because they represent different linear combinations of the same shapes.

We find that the template built from $F_{\rm equil.}$ and $F_{\rm orthog.}$ for estimating the full one-parameter family of shapes in single-field inflation is:
\be
F_{\rm template}(k_1,k_2,k_3,c_s,\tilde c_3)= f_{NL}^{\rm equil.}(c_s,\tilde c_3)F_{\rm equil.}(k_1,k_2,k_3)+f_{NL}^{\rm orthog.}(c_s,\tilde c_3)F_{\rm orthog.}(k_1,k_2,k_3)\ .
\ee
Here $ f_{NL}^{\rm equil.}(c_s,\tilde c_3)$ and $f_{NL}^{\rm orthog.}(c_s,\tilde c_3)$ are chosen by assuming that the expectation values of the estimators for our factorizable templates are given by the Fisher matrix prediction using the 3D scalar  product defined in (\ref{eq:scalar_product}):

\bea\label{eq:relationshipp}
\left( \barr{c}
  \fnlequil(c_s,\tilde c_3)  \\
   \fnlorthog(c_s,\tilde c_3) 
\earr \right)
&=& 
\left( \barr{cc}
  \left( \frac{F_{\dot\pi (\partial\pi)^2} \cdot F_{\rm equil.}}{F_{\rm equil.} \cdot F_{\rm equil.}} \right) &
     \left( \frac{F_{\dot\pi^3} \cdot F_{\rm equil.}}{F_{\rm equil.} \cdot F_{\rm equil.}} \right) \\
  \left( \frac{F_{\dot\pi (\partial\pi)^2} \cdot F_{\rm orthog.}}{F_{\rm orthog.} \cdot F_{\rm orthog.}} \right) &
     \left( \frac{F_{\dot\pi^3} \cdot F_{\rm orthog.}}{F_{\rm orthog.} \cdot F_{\rm orthog.}} \right)
\earr \right)_{f_{\rm NL}\cdot \Delta_\Phi^2=1}
\left( \barr{c}f_{NL}^{\dot\pi(\d_i\pi)^2}(c_s) \\ 
 f_{NL}^{\dot\pi^3}(c_s,\tilde c_3) 
\earr \right)  \nn \\
&=& \left( \barr{cc}
 1.040 & 1.210 \\
 0.1079 & -0.06572
\earr \right)
\left( \barr{c}
 f_{NL}^{\dot\pi(\d_i\pi)^2}(c_s)   \\
  f_{NL}^{\dot\pi^3}(c_s,\tilde c_3) \earr \right)
\eea
where $ f_{NL}^{\dot\pi(\d_i\pi)^2}(c_s,\tilde c_3) $ and $f_{NL}^{\dot\pi^3}(c_s,\tilde c_3)$ are given by eq.~(\ref{eq:newfnl}). This choice is dictated by the definition for the estimators for $\fnlequil$ and $\fnlorthog$ that we will make in the next section. See App.~\ref{app:other_definition} for a  more straightforward but still completely equivalent alternative definition.

\begin{figure}[htp]
\centering
\includegraphics[width=8.2cm]{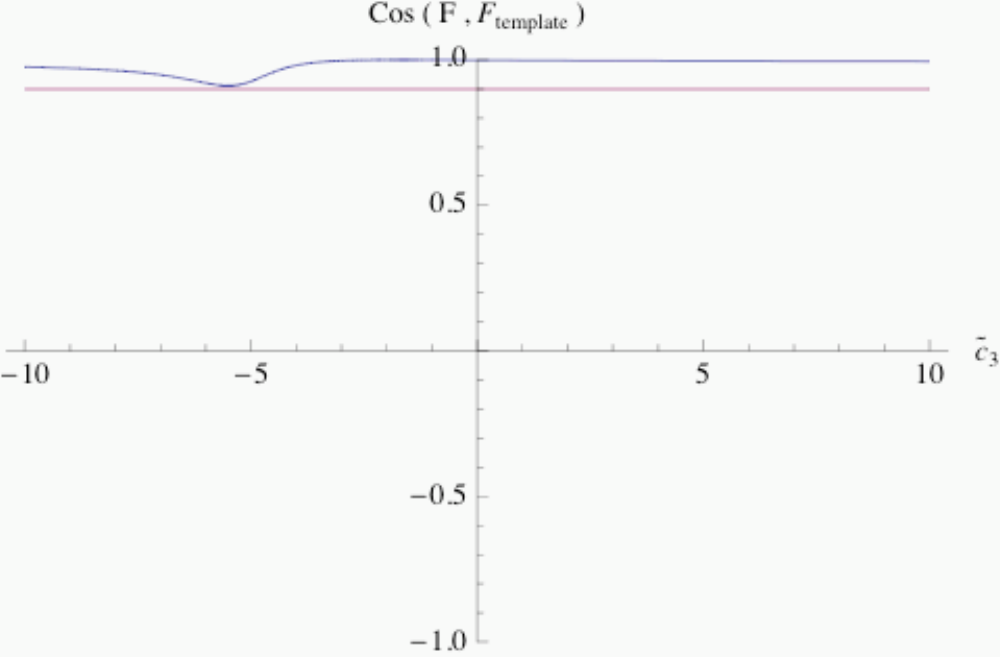}
\caption{\small The cosine between the template shape $F_{\rm template}$ and the exact single-field shape $F$ as we vary $\tilde c_3$, for $c_s\ll1$ where it is independent of $c_s$. The cosine is always very close to one, reaching its minimum, equal to approximately 0.91, for $\tilde c_3\simeq-5.5$. To help visualization, we plot also the line corresponding to a cosine equal to 0.9.}
\label{fig:scalar_template}
\end{figure}

\begin{figure}[htp]
\centering
\includegraphics[width=8.2cm]{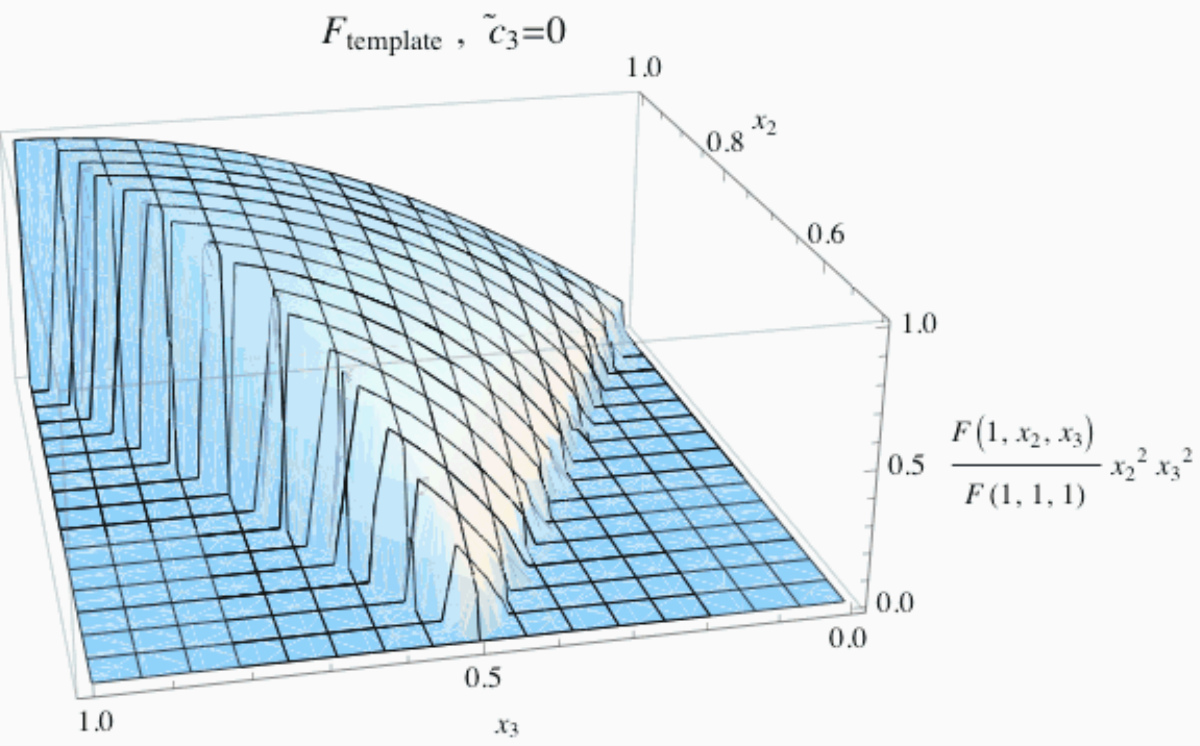}
\includegraphics[width=8.2cm]{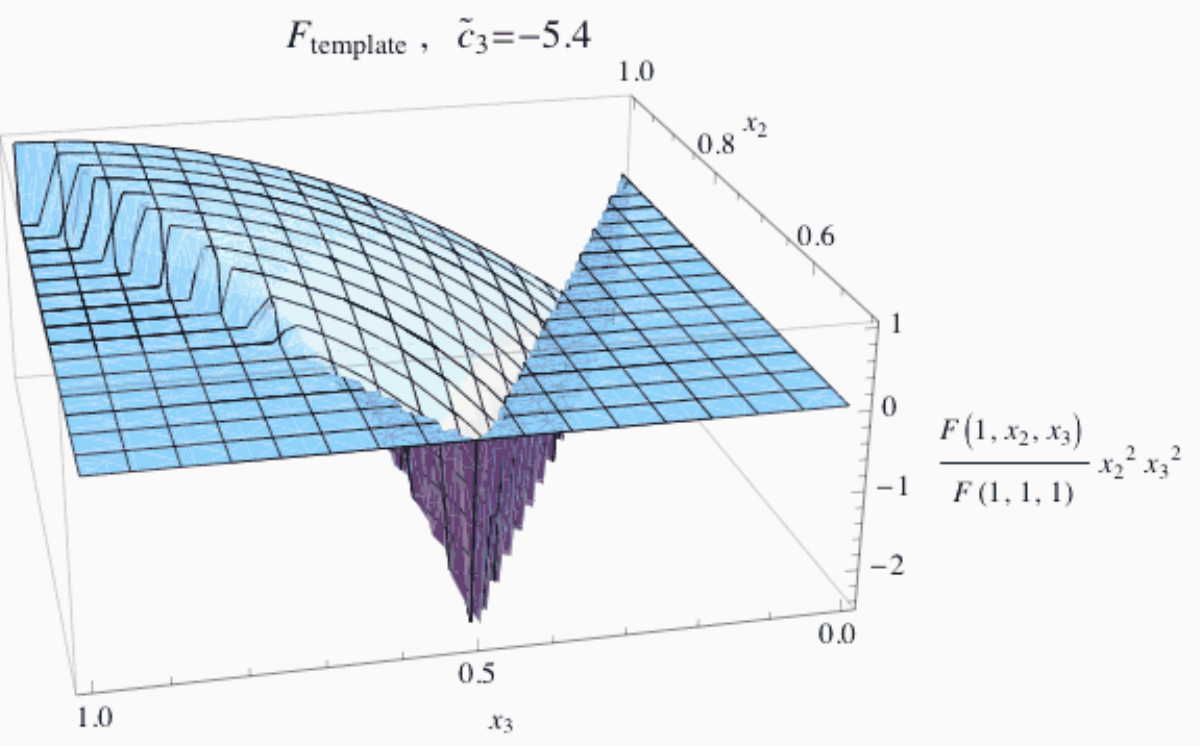}
\caption{\small {\it Left:} $F_{\rm template}$ for $\tilde c_3=0$. This plot should be compared with the left-top panel of Fig.~(\ref{fig:shape}).  {\it Right:} $F_{\rm template}$ for $\tilde c_3=-5$. This plot should be compared with the right-top panel of Fig.~(\ref{fig:shape}).  }
\label{fig:template_shape}
\end{figure}

\begin{figure}[htp]
\centering
\includegraphics[width=8.2cm]{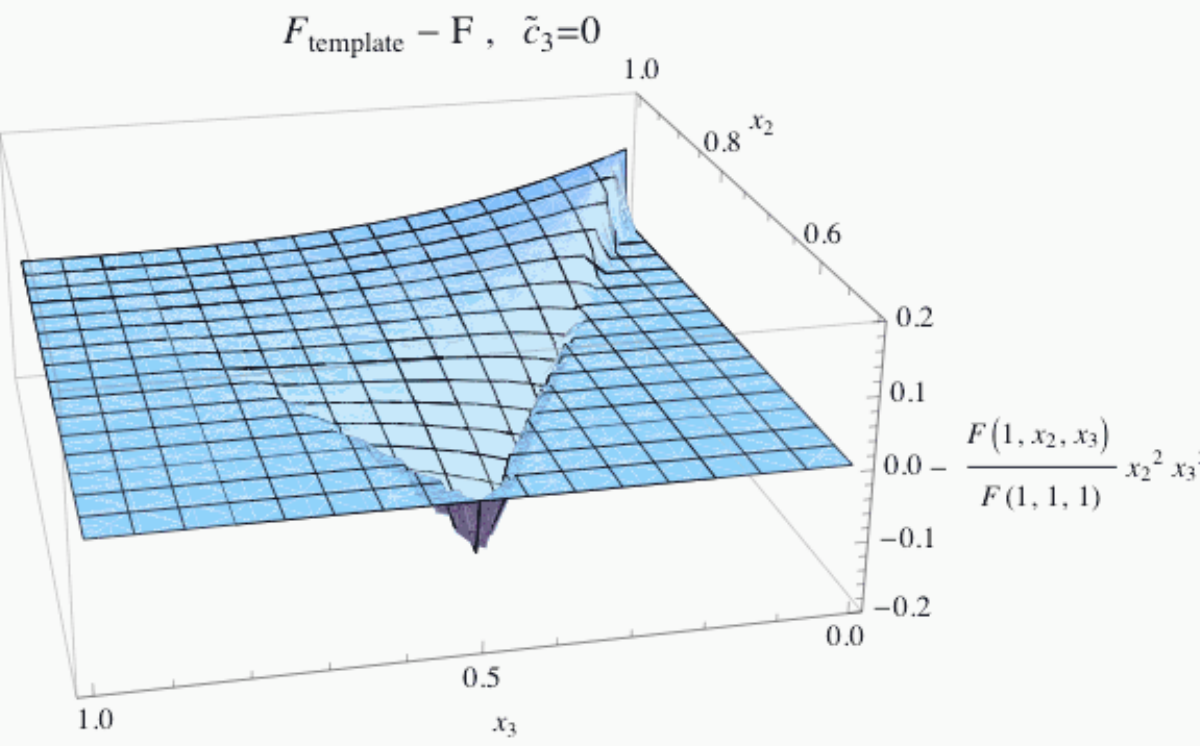}
\includegraphics[width=8.2cm]{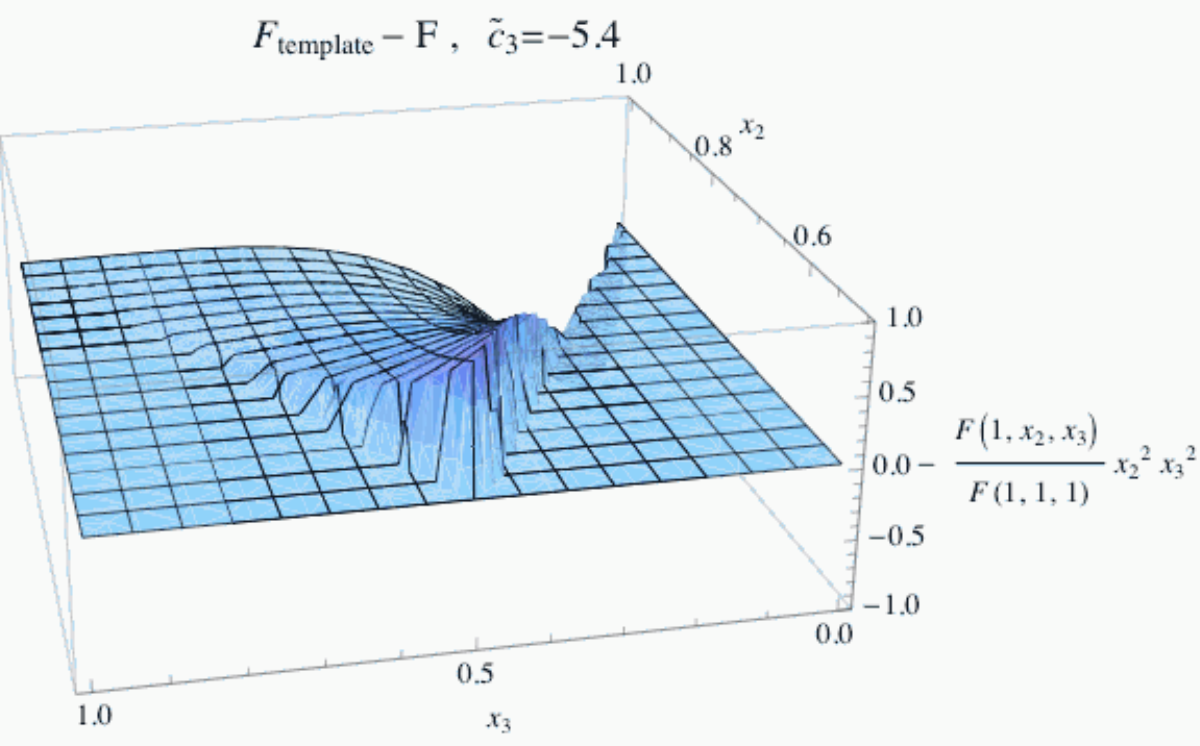}
\caption{\small Difference between the template shape $F_{\rm template}$ and the exact shape $F$ for $\tilde c_3=0$ (on the left) and for $\tilde c_3=-5$ (on the right). Notice that the two figures have different scales on the vertical axes.}
\label{fig:shape-difference}
\end{figure}

  
Quite remarkably, the cosine of this template with the exact shape for any value of $c_s$ and $\tilde c_3$ is always larger than $0.91$. Notice  that this cosine is defined using the $\fnlequil$ and $\fnlorthog$ from eq.~(\ref{eq:relationship}) of App.~\ref{app:other_definition}. This is in fact the definition of $\fnlequil$ and $\fnlorthog$ that minimizes the norm of the shape $F-F_{\rm template}$ with respect to the 3D scalar product. This is in fact the choice that makes the template shape the best approximation to the one of single field inflation for any value of $c_s$ and $\tilde c_3$~\footnote{While this choice of $\fnlequil$ and $\fnlorthog$ is better for this kind of comparison, it is just a linear transformation of the $f_{NL}$'s defined in eq.~(\ref{eq:relationshipp}) that affects neither the minimum of the cosine, nor, as shown in App.~\ref{app:other_definition}, the constraints on the parameters of the Lagrangian.}. The cosine is shown in Fig.~\ref{fig:scalar_template}, where we see that it is always very close to one. This tells us that our two independent shapes are covering the parameter space of the shapes generated in single-field inflation quite well. The scalar product with the exact single-field shape is minimal  when $\tilde c_3\simeq -5.4$, where it is approximately equal to $0.91$. This value of $\tilde c_3$ corresponds to when the shape of the exact single-field non-Gaussianities approaches the orthogonal shape (see upper-right panel of Fig.~\ref{fig:shape}). This means that our template of the orthogonal shape is not extremely accurate, though it is still a very good approximation. Errors up to $9\%$ are satisfactory in the absence of a detection of a non-Gaussian signal. The main reason why our approximation is not better is that the  template $F_{\rm orthog.}(k_1,k_2,k_3)$ does not have the correct behavior in the squeezed limit where $k_1\ll k_2,k_3$. In this regime, the single-field shape $F(k_1,k_2,k_3)$ is of order ${\cal{O}}(1/k_1)$, while our orthogonal template is of order ${\cal{O}}(1/k_1^2)$ (see Fig.~\ref{fig:template_shape} and~\ref{fig:shape-difference}, where again we use the $\fnlequil$ and $\fnlorthog$ from eq.~(\ref{eq:relationship})). This is not a major problem, as what is the relevant number for the similarity of the shapes is the value of the cosine. In App.~\ref{App:template} we show a simple generalization of the template we use, which has a cosine with the exact shape always larger that $0.99$. Though clearly extremely good, this generalized template is rather difficult to implement numerically and therefore given that our template is already a good approximation to $F$ in all the parameter space and that we will have no detection of non-Gaussianity we believe our analysis is sufficient. 

 By inverting the above relationship (\ref{eq:relationshipp}), we obtain the estimate for $c_s$ and $\tilde c_3$ in terms of $f_{NL}^{\rm equil.}$ and $f_{NL}^{\rm orthog.}$.
These expressions will allow us to  translate the limits on the $f_{\rm NL}$ parameters into limits directly on the parameters of the Goldstone boson Lagrangian~(\ref{eq:Lagrangian}) ~\footnote{In \cite{Meerburg:2009ys}, it was argued that the shape of the non-Gaussianity induced by the initial state of each mode as not being in the standard Bunch-Davies vacuum is  well approximated by what the authors of \cite{Meerburg:2009ys} call the Enfolded template. Though we find the theoretical motivation of this effect quite unclear, our parameter space is so general that it includes the enfolded shape, which is given by the following choice of the $f_{NL}$ parameters: 
\be 
 f_{NL}^{\rm equil.}= \frac{1}{12}\; f_{NL}^{\rm enf.} \ , \qquad f_{NL}^{\rm orthog.}= -\frac{1}{12}\; f_{NL}^{\rm enf.}\ ,
\ee
where we have defined $f_{NL}^{\rm enf.}\Delta_{\Phi}^2$ to be the coefficient that multiplies the enfolded shape presented in   \cite{Meerburg:2009ys}.}.

\subsection{Near-de-Sitter limit}

When the inflationary spacetime happens to be very close to de Sitter, the inequality in~(\ref{eq:ds_limit}) can be satisfied, and the behavior of the system changes. In sec.~\ref{sec:EFTds}, we have seen than there are two possible regimes: one in which the goldstone boson has  a linear dispersion relation $\omega=c_s k$, with a small speed of sound \cite{Cheung:2007st,Cheung:2007sv,Creminelli:2006xe}, and one in which the dispersion relation is quadratic in the wavenumber $\omega\propto k^2$. This last regime is known as Ghost Inflation \cite{Cheung:2007st,Cheung:2007sv,ArkaniHamed:2003uz,Senatore:2004rj}. If $d_1\gtrsim4( d_2+d_3)^{1/2}$, then the operator which induces a linear dispersion relation dominates. In this case, we have seen that there are three operators which can induce large non-Gaussianities. The resulting non-Gaussianity is dependent on just two combinations of coefficients: the speed of sound $c_s$ (equivalent to the parameter $d_1$) and the coefficient $\tilde c_3/(d_2+d_3)^4$. The same discussion of the former section applies also in this case: the shape induced by each of the operators is peaked on equilateral configurations, but the total signal can in principle be peaked on even flat-triangle configuration. This requires a more complete analysis that is able to constrain the signal peaked on flat-triangles. To this purpose, the two templates we defined in the former section are enough~\footnote{Notice that for the particular sub-case we are talking about, since the shape $F_{(\d_i\pi)(\d_j\pi)^2}$ is a linear combination of the other two shapes $F_{\dot\pi(\d_i\pi)^2}$ and $F_{\dot\pi^3}$, we can use the result of the former section and conclude that the cosine of our template with the one produced by these models of inflation will  be larger than 0.91. }, and we just need to give the Fisher matrix for this case as well:
\bea\label{eq:relationshipdscs}
&&\!\!\!\!\!\!\!\!\!\left( \barr{c}
  \fnlequil(d_1,\tilde c_3)  \\
   \fnlorthog(d_1,\tilde c_3) 
\earr \right)
= \\
&&\!\!\!\!\!\!\!\!\!\left( \barr{ccc}
  \left( \frac{F_{\dot\pi (\partial\pi)^2} \cdot F_{\rm equil.}}{F_{\rm equil.} \cdot F_{\rm equil.}} \right) &
     \left( \frac{F_{\dot\pi^3} \cdot F_{\rm equil.}}{F_{\rm equil.} \cdot F_{\rm equil.}}\right) &  \left(\frac{F_{(\d_j^2\pi) (\partial\pi)^2} \cdot F_{\rm equil.}}{F_{\rm equil.} \cdot F_{\rm equil.}} \right) \\
  \left( \frac{F_{\dot\pi (\partial\pi)^2} \cdot F_{\rm orthog.}}{F_{\rm orthog.} \cdot F_{\rm orthog.}} \right) &
     \left( \frac{F_{\dot\pi^3} \cdot F_{\rm orthog.}}{F_{\rm orthog.} \cdot F_{\rm orthog.}}\right) &  \left(\frac{F_{(\d_j^2\pi) (\partial\pi)^2} \cdot F_{\rm orthog.}}{F_{\rm orthog.} \cdot F_{\rm orthog.}}   \right)
\earr \right)_{f_{\rm NL}\cdot \Delta_\Phi^2=1}
\left( \barr{c}  \vspace{0.1cm}
f_{NL}^{\dot\pi(\d_i\pi)^2}(d_1) \\ \vspace{0.1cm}
 f_{NL}^{\dot\pi^3}(\tilde c_3/(d_2+d_3)^4) \\
 f_{NL}^{(\d_j^2\pi)(\d_i\pi)^2}(d_1) 
\earr \right)  \nn \\ \nonumber
&&\!\!\!\!\!\!\!\!\!= \left( \barr{ccc}
 1.040 & 1.210 & 0.9878\\
 0.1079 & -0.06572 & 0.1613 
\earr \right)
\left( \barr{c} \vspace{0.1cm}
 f_{NL}^{\dot\pi(\d_i\pi)^2}(d_1)   \\ \vspace{0.1cm}
  f_{NL}^{\dot\pi^3}(\tilde c_3/(d_2+d_3)^4) \\ 
  f_{NL}^{(\d_j^2\pi)(\d_i\pi)^2}(d_1) \earr \right)\ .
\eea
This relationship is valid when the inequality (\ref{eq:ds_limit}) is satisfied and $d_1\gtrsim4( d_2+d_3)^{1/2}$. Notice that since there are two independent parameters, $d_1$ and $\tilde c_3/(d_2+d_3)^4$, the non-Guassianity in this case can be written as a linear combination of two shapes, each one proportional to $1/d_1^{8/5}$ and $\tilde c_3/(d_2+d_3)^4$. The resulting Fisher matrix can be simply obtained from the one above after using eq.~(\ref{eq:newfnlcs}), and it reads
\bea\label{eq:relationshipdscs2}
&&\left( \barr{c}
  \fnlequil(d_1,\tilde c_3)  \\
   \fnlorthog(d_1,\tilde c_3) 
\earr \right)
= \left( \barr{cc}
  -6.791\cdot 10^3 & -4.979\cdot 10^{-2}   \\
 -9.441 \cdot 10^2 & 2.705 \cdot 10^{-3}
 \earr \right)
\left( \barr{c}  \vspace{0.1cm}
\frac{1}{d_1^{8/5}} \\ \vspace{0.1cm}
\frac{\tilde c_3}{(d_2+d_3)^4} 
\earr \right)  
\eea

Still close to de Sitter, but in the opposite limit where $|d_1|\lesssim4( d_2+d_3)^{1/2}$, the system approaches Ghost inflation. In this case the same discussion of the case  $d_1\gtrsim4( d_2+d_3)^{1/2}$ applies, with the only difference that now the operator $\dot\pi^3$ is always irrelevant. In particular the two templates we use are still a good approximation: the cosine never goes below 0.9. The resulting Fisher matrix is given by
\bea\label{eq:relationshipdsghost}
&&\left( \barr{c}
  \fnlequil(d_1,d_2+d_3)  \\
   \fnlorthog(d_1,d_2+d_3) 
\earr \right)
=\\ \nonumber 
&&\left( \barr{cc}
  \left( \frac{\bar F_{\dot\pi (\partial\pi)^2} \cdot F_{\rm equil.}}{F_{\rm equil.} \cdot F_{\rm equil.}} \right) &
     \left(\frac{\bar F_{(\d_j^2\pi) (\partial\pi)^2} \cdot F_{\rm equil.}}{F_{\rm equil.} \cdot F_{\rm equil.}} \right) \\
  \left( \frac{\bar F_{\dot\pi (\partial\pi)^2} \cdot F_{\rm orthog.}}{F_{\rm orthog.} \cdot F_{\rm orthog.}} \right) &
\left(\frac{\bar  F_{(\d_j^2\pi) (\partial\pi)^2} \cdot F_{\rm orthog.}}{F_{\rm orthog.} \cdot F_{\rm orthog.}}   \right)
\earr \right)_{f_{\rm NL}\cdot \Delta_\Phi^2=1}
\left( \barr{c}f_{NL}^{\dot\pi(\d_i\pi)^2}(d_2+d_3) \\ 
 f_{NL}^{(\d_j^2\pi)(\d_i\pi)^2}(d_1,d_2+d_3) 
\earr \right)  \nn \\ \nonumber
&&= \left( \barr{cc}
 0.8625 &  0.9685\\
 0.2621 & 0.1667 
\earr \right)
\left( \barr{c}
 f_{NL}^{\dot\pi(\d_i\pi)^2}(d_2+d_3)    \\
  f_{NL}^{(\d_j^2\pi)(\d_i\pi)^2}(d_1,d_2+d_3) \earr \right)\ .
\eea
Notice that the bar over $F_{\dot\pi (\partial\pi)^2}$ and $F_{(\d_j^2\pi) (\partial\pi)^2}$ means that those are the shapes generated in Ghost  inflation respectively by the operators $\dot\pi (\partial\pi)^2$ and $(\d_j^2\pi) (\partial\pi)^2$. Because of the peculiar dispersion relation $\omega\propto k^2$, the form of this shapes is different from the ones of eq.~(\ref{eq:shapeone}) and~(\ref{eq:Fhigherderiv}), and can only be computed numerically as in eq.~(\ref{eq:ghost1}) and~~(\ref{eq:ghost2}).  We remind the reader that the above relationship is valid when the inequality (\ref{eq:ds_limit}) is satisfied and $|d_1|\lesssim 4(d_2+d_3)^{1/2}$.

\subsection{Negative $c_s^2$} 

In sec.~\ref{sec:negatice_cssq} we saw that still quite close to the de Sitter limit, when $\dot H>0$ or when $d_1<0$, the squared speed of sound $c_s^2$ of the fluctuations at horizon crossing can be negative. In this case the non-Gaussianities tend to be exponentially large. In these models  the inequality (\ref{eq:ds_limit}) can be violated, but not too much, as the operator in $k^4$ has to dominate quite early at high energies in order not to have an exponentially large level of non-Gaussianities. In this sense, also these models are quite close to de Sitter space.

In the case in which $d_1$ is negative, with $d_1<-4(d_2+d_3)^2$ and  $|d_1|\gtrsim 4|\dot H| \mpl^2/(M^3 H)$, then the operator that dominates at horizon crossing is the one proportional to $d_1$. In this case, there are two shapes for the non-Gaussianities that can be large, and the Fisher matrix relation with our templates is given by:
\bea\label{eq:relationshipdscsnegative1}
&&\left( \barr{c}
  \fnlequil(d_1,|c_s|)  \\
   \fnlorthog(d_1,|c_s|) 
\earr \right)
= \\
&&\left( \barr{cc}
  \left( \frac{F_{\dot\pi (\partial\pi)^2} \cdot F_{\rm equil.}}{F_{\rm equil.} \cdot F_{\rm equil.}} \right) &
    \left(\frac{F_{(\d_j^2\pi) (\partial\pi)^2} \cdot F_{\rm equil.}}{F_{\rm equil.} \cdot F_{\rm equil.}} \right) \\
  \left( \frac{F_{\dot\pi (\partial\pi)^2} \cdot F_{\rm orthog.}}{F_{\rm orthog.} \cdot F_{\rm orthog.}} \right) &
\left(\frac{F_{(\d_j^2\pi) (\partial\pi)^2} \cdot F_{\rm orthog.}}{F_{\rm orthog.} \cdot F_{\rm orthog.}}   \right)
\earr \right)_{f_{\rm NL}\cdot \Delta_\Phi^2=1}
\left( \barr{c}  \vspace{0.1cm}
f_{NL}^{\dot\pi(\d_i\pi)^2}(d_1,|c_s|) \\ \vspace{0.1cm}
 f_{NL}^{(\d_j^2\pi)(\d_i\pi)^2}(d_1,|c_s) 
\earr \right)  \nn \\ \nonumber
&&= \left( \barr{cc}
 1.040  & 0.9878\\
 0.1079  & 0.1613 
\earr \right)
\left( \barr{c}  \vspace{0.1cm}
f_{NL}^{\dot\pi(\d_i\pi)^2}(d_1,|c_s|) \\ \vspace{0.1cm}
 f_{NL}^{(\d_j^2\pi)(\d_i\pi)^2}(d_1,|c_s|) 
\earr \right) \ ,
\eea
where $f_{NL}^{\dot\pi(\d_i\pi)^2}(d_1,|c_s|) $ and $ f_{NL}^{(\d_j^2\pi)(\d_i\pi)^2}(d_1,|c_s|) $ are given by eq.~(\ref{eq:fnlcs_negative_1}). It turns out that the correlation of the non-Gaussianity generated in this particular model with the equilateral template is always very large (the cosine is greater than 0.999). For this reason, we will perform the analysis for this model using only the equilateral template.

The other case in which it is possible to have $c_s^2<0$ is when $\dot H>0$, $|d_1|\lesssim 4\dot H \mpl^2/(M^3 H)$, and  $(d_2+d_3)\lesssim 8 M^6/(H^2\dot H \mpl^2) $. In this case, the kinetic operator that dominates at horizon crossing is the one proportional to $\dot H$, and the Fisher matrix is equal to the one above, with the replacement of the $f_{NL}$'s with the one of eq.~(\ref{eq:fnlcs_negative_2}). Notice that this inflationary model predicts a blue tilt of gravity waves \cite{Senatore:2004rj,Creminelli:2006xe}.

Summarizing, Eqs.~(\ref{eq:relationshipp}), (\ref{eq:relationshipdscs}), (\ref{eq:relationshipdsghost}) and (\ref{eq:relationshipdscsnegative1}) are what is necessary to translate the limits we will obtain on the templates $F_{\rm equil.}$ and $F_{\rm orthog.}$ into limits on the parameters of the Lagrangian of single field inflation (\ref{eq:Lagrangian}), as we are now going to do.
  
\section{Results from WMAP\label{sec:analysis}}

\subsection{Analysis pipeline}

We use the pipeline from \cite{Smith:2009jr} to obtain constraints on $\fnlequil$ and $\fnlorthog$ from the 5-year WMAP data.
Since the relevant bispectra are factorizable in $k_1,k_2,k_3$, the optimal estimators can be written down in a straightforward way following 
\cite{Wang:1999vf,Creminelli:2005hu,Komatsu:2003iq}.
We define the functions:
\bea
\alpha_\ell(r) &=& \int \frac{2k^2\, dk}{\pi} \Delta_\ell^T(k) j_\ell(kr)  \\
\beta_\ell(r) &=& \int \frac{2k^2\, dk}{\pi} \Delta_\ell^T(k) P(k) j_\ell(kr)  \\
\gamma_\ell(r) &=& \int \frac{2k^2\, dk}{\pi} \Delta_\ell^T(k) P(k)^{1/3} j_\ell(kr)  \\
\delta_\ell(r) &=& \int \frac{2k^2\, dk}{\pi} \Delta_\ell^T(k) P(k)^{2/3} j_\ell(kr)
\eea
where $P(k)$ is the initial power spectrum of the Newtonian potential $\Phi$ and $\Delta_\ell^T(k)$ is the temperature transfer function.

We analyze the six five-year WMAP maps in V+W channels at Healpix resolution $N_{\rm side}=512$, using the kq75 mask described in \cite{Gold:2008kp}.
These maps are reduced to a single harmonic-space map $(C^{-1}a)_{\ell m}$ as described in Appendix~A of \cite{Smith:2009jr}.
The inverse (signal + noise) filter $C^{-1} = (S+N)^{-1}$ includes the sky cut, inhomogenous noise, and channel-dependent beam
in the definition of the noise covariance.
We also marginalize the monopole, dipole and foreground templates for synchrotron, free-free and dust independently in each of the six channels.

For each value of $r$, we define pixel-space maps $A,B,C,D$ by filtering the WMAP data $(C^{-1}a)$ as follows:
\bea
A(\n,r) &=& \sum_{\ell m} \alpha_\ell(r) (C^{-1}a)_{\ell m} Y_{\ell m}(\n) \\
B(\n,r) &=& \sum_{\ell m} \beta_\ell(r) (C^{-1}a)_{\ell m} Y_{\ell m}(\n) \\
C(\n,r) &=& \sum_{\ell m} \gamma_\ell(r) (C^{-1}a)_{\ell m} Y_{\ell m}(\n) \\
D(\n,r) &=& \sum_{\ell m} \delta_\ell(r) (C^{-1}a)_{\ell m} Y_{\ell m}(\n)
\eea
The optimal estimators $\hfnlequil$ and $\hfnlorthog$ are then given by:
\bea
\hfnlequil &=& \frac{1}{N_{\rm equil}} \int dr\, r^2 \Big[ -3 A(\n,r) B(\n,r)^2 + 6 B(\n,r) C(\n,r) D(\n,r) - 2 D(\n,r)^3  \nn \\
&&
+6 \left\langle A(\n,r) B(\n,r)\right\rangle_{\rm MC} B(\n,r)
+3 \left\langle B(\n,r) B(\n,r)\right\rangle_{\rm MC} A(\n,r) \nn \\
&&
-6 \left\langle B(\n,r) C(\n,r)\right\rangle_{\rm MC} D(\n,r)
-6 \left\langle B(\n,r) D(\n,r)\right\rangle_{\rm MC} C(\n,r) \nn \\
&&
-6 \left\langle B(\n,r) C(\n,r)\right\rangle_{\rm MC} D(\n,r)
+6 \left\langle D(\n,r) D(\n,r)\right\rangle_{\rm MC} D(\n,r)
\Big]  \label{eq:equil_estimator}  \\
\hfnlorthog &=& \frac{1}{N_{\rm orthog}} \int dr\, r^2  \Big[ -9 A(\n,r) B(\n,r)^2 + 18 B(\n,r) C(\n,r) D(\n,r) - 8 D(\n,r)^3  \nn \\
&&
+18 \left\langle A(\n,r) B(\n,r)\right\rangle_{\rm MC} B(\n,r)
+9 \left\langle B(\n,r) B(\n,r)\right\rangle_{\rm MC} A(\n,r) \nn \\
&&
-18 \left\langle B(\n,r) C(\n,r)\right\rangle_{\rm MC} D(\n,r)
-18 \left\langle B(\n,r) D(\n,r)\right\rangle_{\rm MC} C(\n,r) \nn \\
&&
-18 \left\langle B(\n,r) C(\n,r)\right\rangle_{\rm MC} D(\n,r)
+24 \left\langle D(\n,r) D(\n,r)\right\rangle_{\rm MC} D(\n,r)
\Big]  \label{eq:orthog_estimator}
\eea
Each estimator contains a 3-point term and a 1-point term which improves the variance of the estimator in the
presence of inhomogeneous noise or a sky cut, as described in \cite{Creminelli:2005hu}.
Whenever an expression such as $\langle A(\n,r) B(\n,r) \rangle_{\rm MC}$ appears in Eqs.~(\ref{eq:equil_estimator}),~(\ref{eq:orthog_estimator}),
it denotes a Monte Carlo average in which we construct the filtered maps ($A(\n,r)$ and $B(\n,r)$ in this case) from signal + noise simulations instead
of the WMAP data, and take the average over many simulations.

In implementation, the $r$ integrals in the definitions of $\hfnlequil,\hfnlorthog$ must be replaced by finite sums.
We use the optimization algorithm from \cite{Smith:2006ud} to minimize the number of sampling points, finding that 30 and 54
$r$-values are necessary, for the equilateral and orthogonal shapes respectively, to preserve the Fisher matrix to
one part in $10^6$.

We compute the normalization constants $N_{\rm equil}, N_{\rm orthog}$ by running the pipeline on ensembles of non-Gaussian
simulations, using the simulation algorithm from \cite{Smith:2006ud}.
This ensures that each estimator is an unbiased estimator of the corresponding $f_{NL}$ parameter, {\em assuming} no additional
contributions to the bispectrum~\footnote{Some secondary contributions to $\hat f_{NL}$ have been studied and have all been predicted to be small compared to the WMAP statistical error
\cite{Smith:2009jr,Smith:2006ud,Seljak:1998nu,Cooray:1999kg,Goldberg:1999xm,Verde:2002mu,Castro:2002df,Bartolo:2006cu,Babich:2008uw,Pitrou:2008ak,Senatore:2008vi,Senatore:2008wk,Khatri:2008kb,Nitta:2009jp},
although it is still not clear that all possible important secondaries have been studied.
}:
\bea
\left\langle \hfnlequil \right\rangle &=& \fnlequil \qquad\mbox{(assuming $\fnlloc=\fnlorthog=\cdots=0$)}  \nn \\
\left\langle \hfnlorthog \right\rangle &=& \fnlorthog \qquad\mbox{(assuming $\fnlloc=\fnlequil=\cdots=0$)}
\eea
(We have written ``$\cdots$'' here to indicate other possible terms in the 3-point function, for example point sources.)
The covariance of the estimators $\fnlequil,\fnlorthog$ is also determined by Monte Carlo.

There is an alternate version of this construction, in which the estimators $\hfnlequil$, $\hfnlorthog$, $\cdots$
are defined differently (by taking linear combinations) in such a way that each estimator has unit response to the corresponding
$f_{NL}$ parameter and zero response to the other $(N-1)$ parameters.
Details of this construction are given in Appendix~\ref{app:other_definition}.
For now we note in advance that the analysis of single-field inflation in the following subsection will not depend 
on which definition is used; it is simply an arbitrary choice that does not affect parameter constraints.
We have chosen to use the definition in Eqs.~(\ref{eq:equil_estimator}),~(\ref{eq:orthog_estimator}) for consistency with previous 
analyses \cite{Creminelli:2005hu,Creminelli:2006rz,Komatsu:2008hk} so that our limits on $\fnlequil$ in the 
next subsection can be directly compared to results in the literature.

\subsection{WMAP constraints on $\fnlequil, \fnlorthog$}
\label{sec:constraints_error}

We encountered one issue in our analysis which affects previously reported constraints on $\fnlequil$.
In principle, the $r$ integral in Eqs.~(\ref{eq:equil_estimator}),~(\ref{eq:orthog_estimator}) should run from $r=0$ to $r=\infty$.
Previous analyses \cite{Creminelli:2005hu,Creminelli:2006rz} have cut off the integral at the horizon $r_{\rm horiz} \approx 14500$ Mpc.
Empirically, we find that setting $r_{\rm max} \approx r_{\rm horiz} + (300 \mbox{ Mpc})$ is needed for convergence; truncating at the 
horizon significantly underestimates the error $\sigma(\fnlequil)$.

The need to extend the $r$-integral beyond the horizon does not indicate that causality is violated.
The $r$ integral in the estimator arises from writing a delta function $(2\pi)^3 \delta^{3}(k_1+k_2+k_3)$ as an 
integral $\int d^3r e^{ik\cdot r}$, so formal contributions to the estimator from $r > r_{\rm horiz}$ do not correspond
to physical contributions from outside the causal horizon.
However, one can use causality to show that $\alpha_\ell(r)=0$ for $r > r_{\rm horiz}$.
This implies that the estimator for the {\em local} shape can safely be truncated at the horizon, so this issue
does not affect previously reported estimates of $\fnlloc$ (e.g. \cite{Creminelli:2006rz,Yadav:2007yy,Komatsu:2008hk}).

For the five-year WMAP data in (V+W) bands with the optimal estimator, kq75 mask and foreground template marginalization, we find 
$\fnlequil = 155 \pm 140$ and $\fnlorthog = -149 \pm 110$ (errors are $1\sigma$), indicating no detection of non-Gaussianity using either shape. 
The 2-by-2 covariance matrix is given by
\bea\label{eq:covariance}
C &=&
\left( \barr{cc}
    \mbox{Var}(\hfnlequil) & \mbox{Cov}(\hfnlequil,\hfnlorthog)  \\
    \mbox{Cov}(\hfnlequil,\hfnlorthog) & \mbox{Var}(\hfnlorthog)
\earr \right)_{\rm WMAP}  \nn \\
& \approx &
\left( \barr{cc}
  1.96 \times 10^4 & 5.0 \times 10^3 \\
  5.0 \times 10^3 & 1.21 \times 10^4
\earr \right)  \label{eq:estcov}
\eea
where the subscript $_{\rm WMAP}$ reminds us that these quantities depend on the WMAP data.
We find that the correlation $\mbox{Corr}(\hfnlequil, \hfnlorthog)$ in the Monte Carlo simulations is consistent with the
$\approx 0.32$ cross correlation expected from computing the 2D cosine defined in eq.~(\ref{eq:2d_cosine}).

Since our estimators are optimal, the $1\sigma$ error $\sigma(\fnlequil) = 140$ is the best constraint that can be obtained
from the five-year WMAP data, if the $r$-integral in eq.~(\ref{eq:equil_estimator}) is properly taken to $r_{\rm max} = \infty$.
As a check, we verified that the statistical errors agree with the Fisher matrix (assuming $f_{\rm sky}^{-1/2}$ scaling).
For comparison with previous results, we also report constraints on $\fnlequil$ from 1-year and 3-year WMAP data in
Table~\ref{tab:fnl_equilateral}.

\begin{table}
\begin{center}
\begin{tabular}{|c|c|c|}
\hline             & optimal estimator &  suboptimal estimator  \\ \hline
WMAP1 (kp0 mask)  &  $\fnlequil = 125 \pm 177$  &  $\fnlequil = 236 \pm 204$  \\
WMAP3 (kp0 mask) &  $\fnlequil = 146 \pm 149$  &  $\fnlequil = 178\pm 162$  \\
WMAP5 (kq75 mask) &  $\fnlequil = 155 \pm 140$  &  $\fnlequil = 145 \pm 162$  \\ \hline
\end{tabular}
\end{center}
\caption{\small Constraints on $\fnlequil$ from 1-year, 3-year and 5-year WMAP data.
Previously reported results have underestimated the error $\sigma(\fnlequil)$ by truncating the integral in the estimator 
(eq.~(\ref{eq:equil_estimator})) at $r_{\rm max} = r_{\rm horiz} \approx 14500$ Mpc, rather than taking $r_{\rm max}=\infty$.
The results shown here for the optimal estimator represent the best possible statistical error for each dataset.
We note that the statistical errors in WMAP3 and WMAP5 are roughly equal, even though WMAP5 is more sensitive, because the kq75 mask 
introduced in the 5-year analysis is more conservative than the kp0 mask used in the 1-year and 3-year analyses.
For comparison with previous results, we also show the constraint obtained using a suboptimal estimator similar to the
one described in Appendix~A of \cite{Komatsu:2008hk}.}
\label{tab:fnl_equilateral}
\end{table}

\subsection{WMAP constraints: non near-de-Sitter models}
\label{ssec:wmap_single_field}

In the next few subsections, we will use the WMAP measurements $\fnlequil = 155 \pm 140$, $\fnlorthog = -149\pm 110$ to place constraints
on single-field inflation in different model spaces.
In this subsection we will consider the parameter space $(c_s,\tilde c_3)$ of the models not-so-near-de-Sitter described in sec.~\ref{sec:EFT}.
In the following subsections we will analyze the near de Sitter models and the models with $c_s^2<0$.

Our basic tool will be a $\chi^2$ statistic which quantifies agreement between the observed bispectrum and the model bispectrum.
Given model parameters $(c_s,\tilde c_3)$, our estimators $\hfnlequil,\hfnlorthog$ acquire expectation values given by:
\be
\left( \barr{c}
  \langle \hfnlequil(c_s,c_3) \rangle \\
  \langle \hfnlorthog(c_s,c_3) \rangle
\earr \right)
= \left( \barr{cc}
 1.040 & 1.210 \\
 0.1079 & -0.06572
\earr \right)
\left( \barr{c}
   \frac{85}{324} ( 1 - 1/c_s^2 ) \\
   \frac{10}{243} ( 1 - 1/c_s^2 ) (\tilde c_3 + 3 c_s^2 / 2)
\earr \right)  \label{eq:transfer}
\ee
by combining Eqs.~(\ref{eq:newfnl}),~(\ref{eq:relationshipp}) above.

Given estimates $(\hfnlequil)_{\rm WMAP}, (\hfnlorthog)_{\rm WMAP}$ from the WMAP data, with the associated 2-by-2 covariance matrix $C$ of eq.~(\ref{eq:covariance}),
we define a $\chi^2$ statistic by:
\be
\chi^2(c_s,\tilde c_3)_{\rm WMAP} = v(c_s,\tilde c_3)_{\rm WMAP}^T \,\, C^{-1} \, v(c_s,\tilde c_3)_{\rm WMAP}  \label{eq:chi2_def}
\ee
where:
\be
v(c_s,\tilde c_3)_{\rm WMAP} =
  \left( \barr{c}
    \langle\hfnlequil(c_s,\tilde c_3)\rangle - (\hfnlequil)_{\rm WMAP} \\
    \langle\hfnlorthog(c_s,\tilde c_3)\rangle - (\hfnlorthog)_{\rm WMAP}
  \earr \right)
\ee
We have used the notation $\chi^2(c_s,\tilde c_3)_{\rm WMAP}$ to emphasize that the $\chi^2$ statistic depends both on the
model $(c_s,\tilde c_3)$ and the WMAP data, and measures agreement between the two.
If we construct an ensemble of non-Gaussian simulations in a fixed model $(c_s,\tilde c_3)$, and evaluate $\chi^2(c_s,\tilde c_3)_{\rm sim}$
on each simulation using the true $(c_s,\tilde c_3)$ of the model, then it will be distributed as a $\chi^2$ random variable with two degrees of 
freedom (as suggested by the notation).
Turning this around, we can test the hypothesis that a given $(c_s,\tilde c_3)$ is consistent with the WMAP data, by comparing the
value of $\chi^2(c_s,\tilde c_3)_{\rm WMAP}$ to a $\chi^2$ distribution with two degrees of freedom.

\begin{figure}
\centerline{ \includegraphics[width=5.0in]{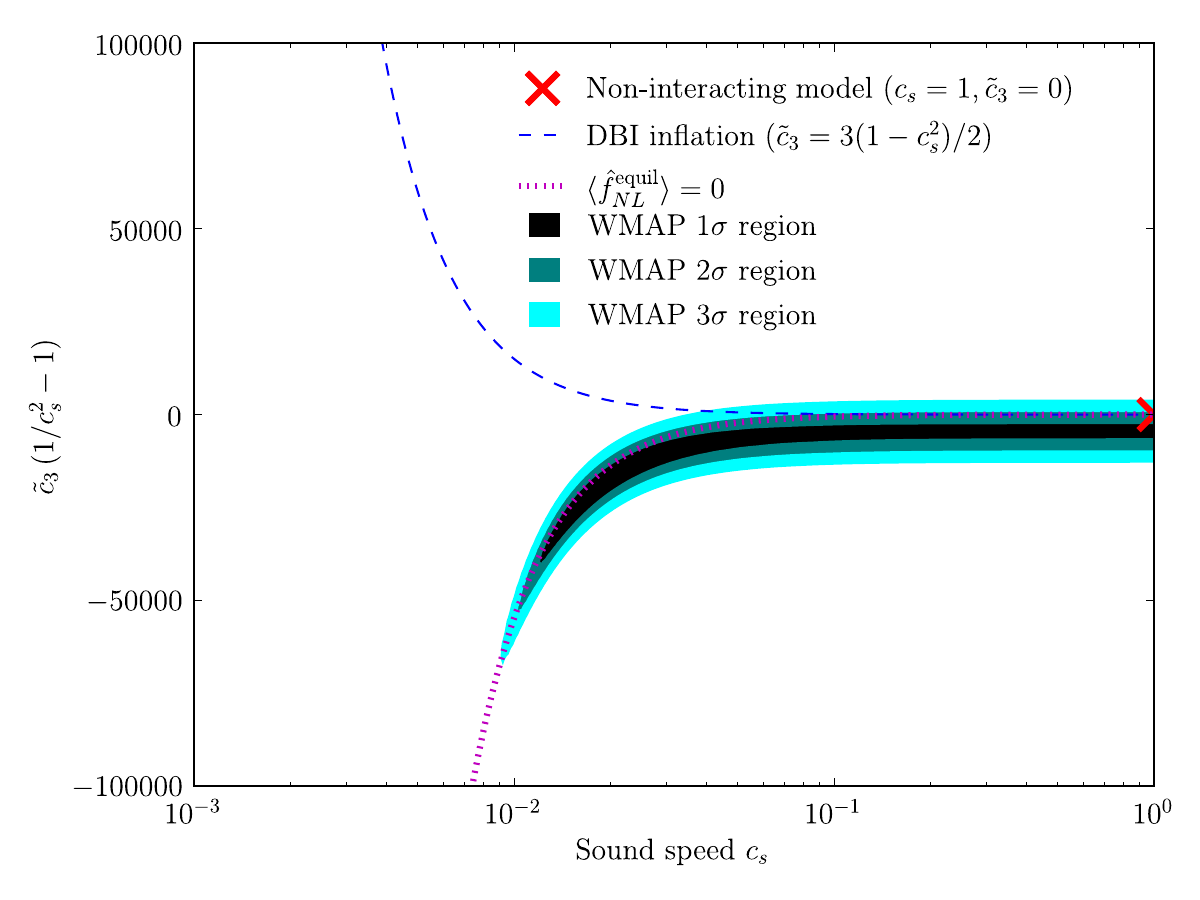} }
\centerline{ \includegraphics[width=5.0in]{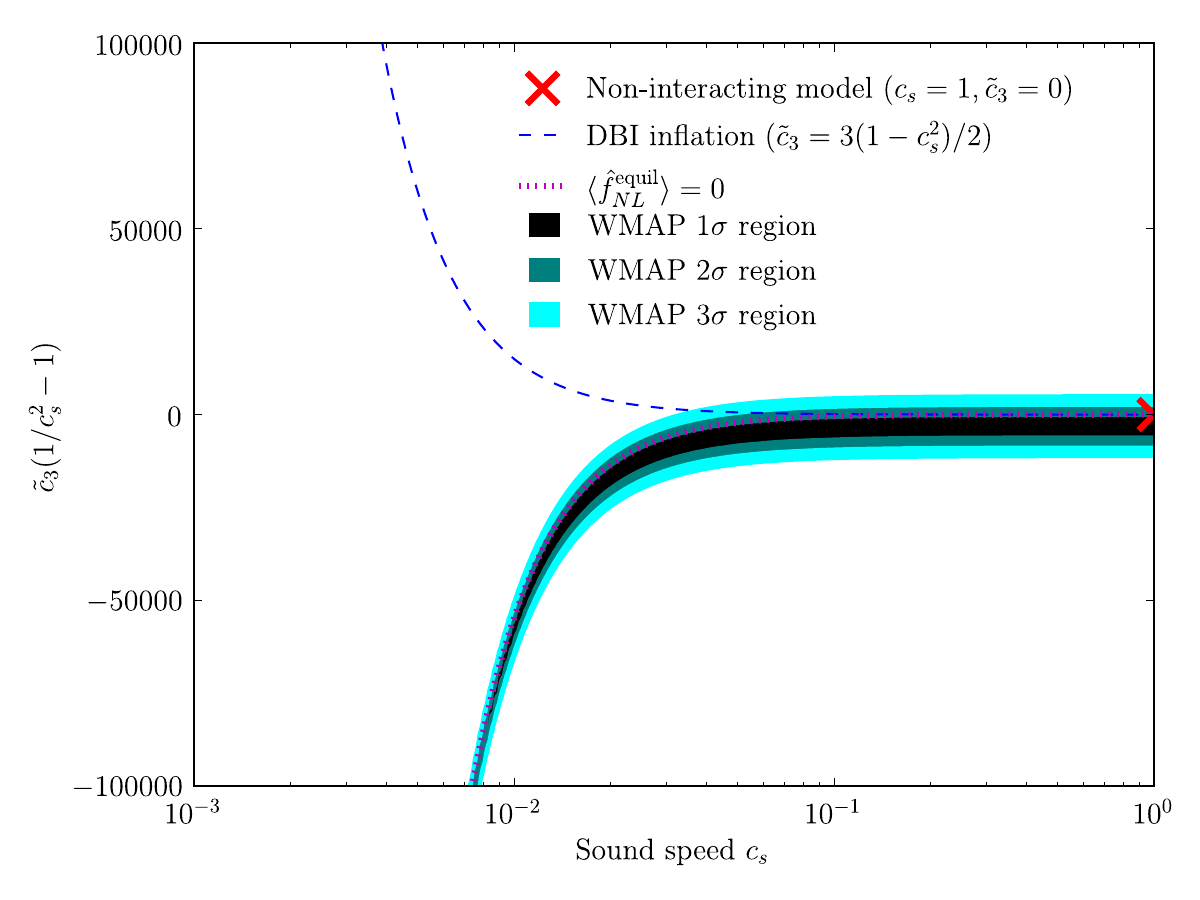} }
\caption{ \small
{\em Top panel:} 68\%, 95\% and 99.7\% confidence regions in the single-field inflation parameters $(c_s,\tilde c_3)$ from five-year WMAP data,
obtained from an analysis which uses $\fnlequil$ and $\fnlorthog$ (eq.~(\ref{eq:2d_pvalue})).
{\em Bottom panel:} Confidence regions obtained from an analysis using $\fnlequil$ alone (eq.~(\ref{eq:1d_chisq})), showing weaker constraints.
}
\label{fig:csc3}
\end{figure}

In Fig.~\ref{fig:csc3}, top panel, we show 68\%, 95\% and 99.7\% confidence regions in the $(c_s,\tilde c_3)$-plane using the WMAP data.
More precisely, we define the confidence region corresponding to $p$-value $p$ (where $p$=0.32, 0.05, or 0.003) by the inequality
\be
\chi^2(c_s,\tilde c_3)_{\rm WMAP} \le F^{-1}(1-p)  \label{eq:2d_pvalue}
\ee
where $F$ is the CDF for a $\chi^2$ random variable with two degrees of
freedom.\footnote{Note that we define confidence regions in the sense of frequentist statistics: we say that a model $(c_s,\tilde c_3)$
lies outside the 95\% confidence region if, in an ensemble of simulations with model parameters $(c_s,\tilde c_3)$, the value of
$\chi^2(c_s,\tilde c_3)_{\rm WMAP}$ is larger than $\chi^2(c_s,\tilde c_3)_{\rm sim}$ 95\% of the time.
The $p$-value that we assign to a model $(c_s,\tilde c_3)$ should be interpreted as the significance of rejecting the null hypothesis that
the model is consistent with the WMAP data, using $\chi^2$ as the test statistic, not as a probability of $(c_s,\tilde c_3)$ being the true 
model.  (For example, if we integrate $p$ over the $(c_s,\tilde c_3)$ plane, the integral will not be equal to one.)
We have used hypothesis testing rather than Bayesian inference for this problem because there is a clear choice of test statistic $\chi^2$,
whereas there would be no clear choice of prior density $P(c_s,\tilde c_3) dc_s d\tilde c_3$ in a Bayesian analysis.}
It is seen that the simplest, non-interacting single-field model ($c_s=1, \tilde c_3=0$) is consistent
with the WMAP data (the $p$-value is $p=0.1$).
The allowed region of parameter space shows a nontrivial structure which arises because the two primordial bispectrum shapes
$F_{\dot\pi(\partial\pi)^2}, F_{\dot\pi^3}$ are nearly degenerate.
There is an allowed region in the lower left corner with $\tilde c_3 \approx -5.4$ and small $c_s$.
In this region, there is a near-cancellation between
the bispectra associated with the operators $\dot\pi(\d_i \pi)^2$ and $\dot\pi^3$ (see Fig.~\ref{fig:scalar_product}).
In the limit $c_s\rightarrow 1$, the parameter $\tilde c_3$ is not bounded, although the WMAP data do give the constraint
$-9920 \le (1-c_s^2)\tilde c_3 \le 1000$ (95\% CL) in this limit.

For comparison, in the bottom panel of Fig.~\ref{fig:csc3}, we show 68\%, 95\% and 99.7\% confidence regions using an analysis
which discards $\fnlorthog$ and obtains constraints from $\fnlequil$ alone.
More precisely, we define an ``equilateral'' $\chi^2$ by:
\be
\chi^2(c_s,\tilde c_3)_{\rm equil} = \frac{(\langle\hfnlequil(c_s,\tilde c_3)\rangle - (\hfnlequil)_{\rm WMAP})^2}{\mbox{Var}(\hfnlequil)}   \label{eq:1d_chisq}
\ee
and define confidence regions by converting to a $p$-value using $\chi^2$ statistics with one degree of freedom.
Using $\fnlequil$ alone, it is seen that there is a degeneracy which allows models with $\tilde c_3 \approx -5.4$ and arbitrarily small $c_s$.
As we move along the degeneracy line in the $c_s \rightarrow 0$ direction, we get large contributions to $\fnlequil$ 
(proportional to $(1-1/c_s^2)$ and $\tilde c_3/c_s^2$) which nearly cancel.
If $\fnlorthog$ is included in the analysis (top panel), the degeneracy is broken for sufficiently small $c_s$,
because a detectably large $\fnlorthog$ is generated.

We can also ask: what is the WMAP constraint on the sound speed $c_s$ during inflation~\footnote{We will constrain $c_s$ now for the models
not-near-de-Sitter, and will come back again to $c_s$ in the next subsection.}?
The answer is different depending on what assumptions we make about the parameter $\tilde c_3$.
If we assume $\tilde c_3=0$, or DBI inflation ($\tilde c_3=3(1-c_s^2)/2$), we can discard
$\fnlorthog$ and constrain $c_s$ using $\fnlequil$ alone.
This is because the bispectrum is always highly correlated to the equilateral shape, so that including
$\fnlorthog$ would not increase the statistical power.
We find the following lower bounds:
\bea
c_s & \gtrsim & 0.048 \qquad\mbox{(95\% CL, $\tilde c_3=0$)}  \\
c_s & \gtrsim & 0.054 \qquad\mbox{(95\% CL, DBI inflation)}
\eea
To get the most conservative constraint on $c_s$, we would marginalize $\tilde c_3$ instead of assuming a specific form.
More precisely, we define a one-variable $\chi^2$ statistic by
evaluating the two-variable $\chi^2(c_s,\tilde c_3)$ at the value of $\tilde c_3$ which
minimizes the $\chi^2$ for a given $c_s$:
\be
\chi^2(c_s)_{\rm marg} = \min_{\tilde c_3} \left( \chi^2(c_s,\tilde c_3)_{\rm WMAP} \right) 
   \qquad\mbox{($\tilde c_3$ marginalized)}  \label{eq:chisq_c3_marginalized}
\ee
Converting this $\chi^2$ to a lower limit using $\chi^2$ statistics with one degree of freedom, we find:
\be
c_s \gtrsim 0.011 \qquad\mbox{($\tilde c_3$ marginalized, 95\% CL)}
\ee
This very general lower limit applies to all single-field inflation models regardless of the value of the coupling $\tilde c_3$.
To obtain it, it is necessary to include the new shape $\fnlorthog$ in the analysis.
Using $\fnlequil$ alone, there is no lower limit on $c_s$ when $\tilde c_3$ is marginalized,
due to the degeneracy seen in the bottom panel of Fig.~\ref{fig:csc3}.

So far in this analysis we have neglected the metric fluctuations. As shown in detail in \cite{Cheung:2007sv}, in these kind of models these give only subleading corrections suppressed by the slow roll parameters, and can be safely neglected.

\subsection{WMAP constraints: near de Sitter models with $d_1 \gtrsim 4(d_2+d_3)^{1/2}$}

We can also use the two-template WMAP analysis to put constraints on the near de Sitter parameter spaces described in sec.~\ref{sec:EFTds}.
In this subsection, we will consider the case $d_1 \gtrsim 4(d_2+d_3)^{1/2}$.

\begin{figure}
\centerline{ \includegraphics[width=5.0in]{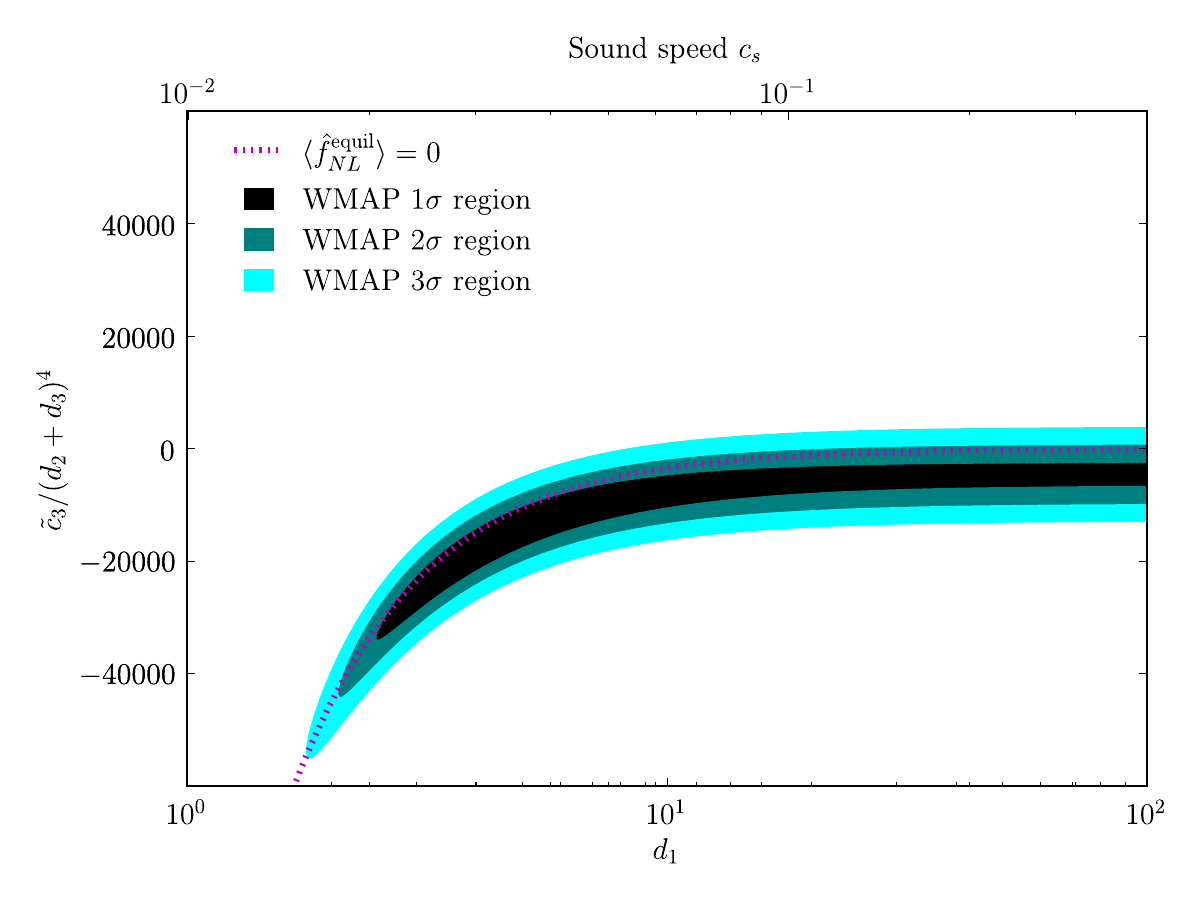} }
\caption{\small WMAP constraints in the $(d_1, {\tilde c_3}(d_2+d_3)^{-4})$-plane, for the near de Sitter model with $d_1 \gtrsim4 (d_2+d_3)^{1/2}$.}
\label{fig:wmap_near_desitter}
\end{figure}

We take the parameters of the model to be $(d_1, {\tilde c_3}(d_2+d_3)^{-4})$.
As in the preceding subsection, we define a $\chi^2$ statistic which quantifies agreement between the model
parameters and the WMAP data, using eq.~(\ref{eq:relationshipdscs2}) above for the expectation values of the $f_{NL}$ estimators.
In detail, $\chi^2$ is defined as a function of $(d_1, {\tilde c_3}(d_2+d_3)^{-4})$ by:
\bea
\left( \barr{c}
  \langle \hfnlequil \rangle \\
  \langle \hfnlorthog \rangle
\earr \right)
&=& \left( \barr{cc}
 -6.791 \times 10^3 & -4.979 \times 10^{-2}   \\
 -9.441 \times 10^2 &  2.705 \times 10^{-3}
\earr \right)
\left( \barr{c}
   d_1^{-8/5} \\
   {\tilde c_3} (d_2+d_3)^{-4}
\earr \right)  \\
v &=&
\left( \barr{c}
    \langle\hfnlequil \rangle - (\hfnlequil)_{\rm WMAP} \\
    \langle\hfnlorthog \rangle - (\hfnlorthog)_{\rm WMAP}
  \earr \right)  \\
\chi^2 &=&
v^T C^{-1} v \qquad\mbox{(where $C=\mbox{Cov}(\hfnlequil,\hfnlorthog)$)}  \label{eq:chisq_near_desitter}
\eea
In Fig.~\ref{fig:wmap_near_desitter}, we show $1\sigma$, $2\sigma$, and $3\sigma$ contours in the $(d_1, {\tilde c_3}(d_2+d_3)^{-4})$ 
plane obtained using this $\chi^2$ statistic.  The non-interacting model ($d_1 \gg 1$, ${\tilde c_3}(d_2+d_3)^{-4}=0$) is consistent
with the data at $2\sigma$.  (This is the case for all parameter spaces considered in this paper, since the $\chi^2$ of the
non-interacting model is independent of the parameter space into which it is embedded.) It is worth noticing that if we imposed $d_2+d_3$ to saturate the bound $d_1 \gtrsim 4(d_2+d_3)^{1/2}$ and we also imposed $\tilde c_3$ not to be too large, then many of the most extreme values of the $y$-axis in Fig.~\ref{fig:wmap_near_desitter} would be excluded.
Using eq.~(\ref{eq:near_desitter_cs}), the sound speed is given by $c_s \approx (0.00993) d_1^{4/5}$, and lower limits on $c_s$ are given by:
\bea
c_s & \gtrsim & 0.073 \qquad\mbox{\ (95\% CL, ${\tilde c_3}(d_2+d_3)^{-4} = 0$)}  \\
c_s & \gtrsim & 0.0189 \qquad\mbox{(95\% CL, ${\tilde c_3}(d_2+d_3)^{-4}$ marginalized)}
\eea
Note that the requirement that $c_s \le 1$ implies $d_1 \lesssim 319$ in this model.

So far we have neglected metric fluctuations. Following \cite{Creminelli:2006xe,Cheung:2007sv} it is straightforward to see that this is a good approximation for $d_1\lesssim (\mpl/(30 M))^{5}$. For the parameter space we are interested in, this translates in $M\lesssim \mpl/10$ to a good approximation. When this inequality is violated, as shown in  \cite{Creminelli:2006xe,Cheung:2007sv}  the squared speed of sound $c_s^2$ of the fluctuations becomes negative and the modes begin to grow exponentially before crossing the horizon. In this case these models become similar to the ones shown in the next sec.~\ref{sec:negativec_s^2}, and, as we will see, are generically very disfavored by the data.

\subsection{Near-de-Sitter models with $|d_1| \lesssim 4(d_2+d_3)^{1/2}$}

\begin{figure}
\centerline{ \includegraphics[width=4.4in]{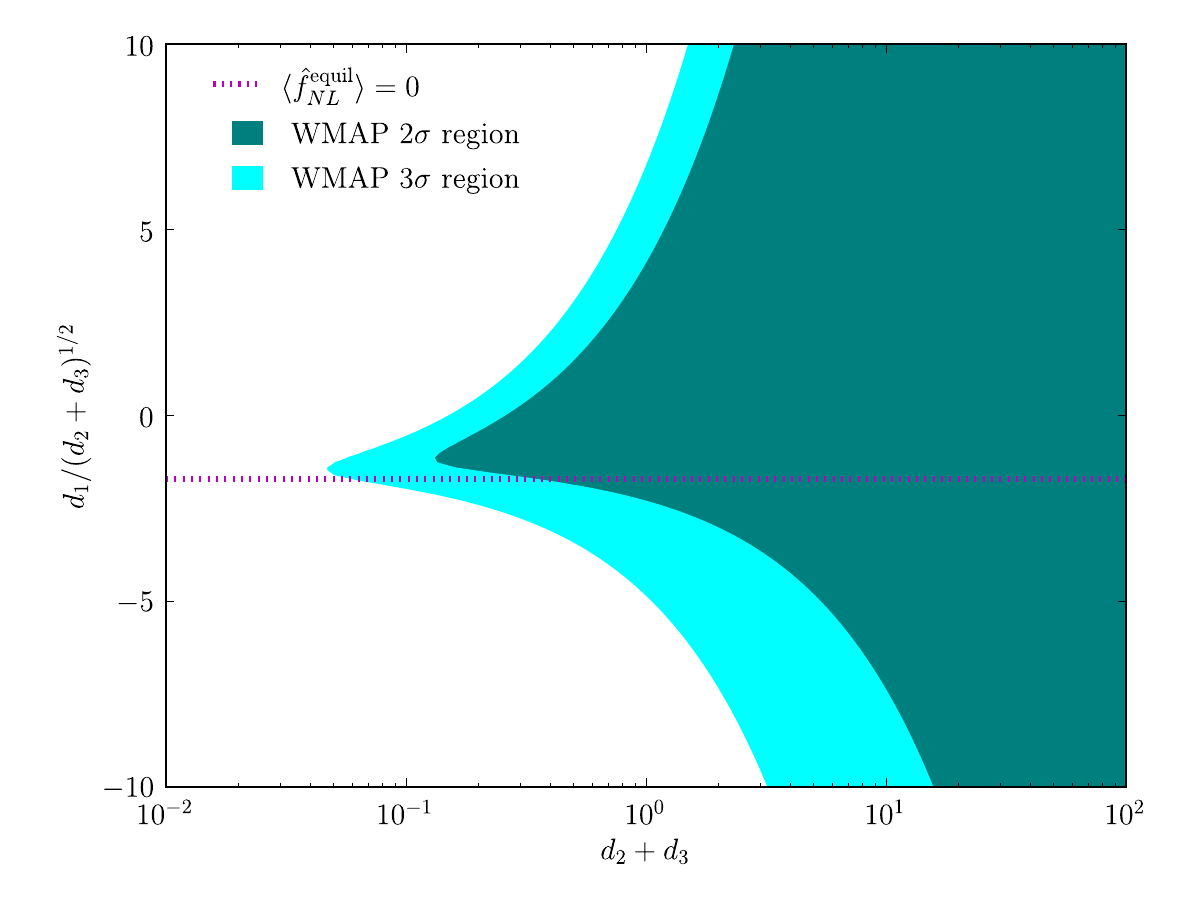} }
\caption{\small WMAP constraints in the $(d_1,d_2+d_3)$ plane, for the near de Sitter model in the ghost inflation limit $|d_1| \lesssim 4(d_2+d_3)^{1/2}$.
Note that the range on the $y$-axis has been chosen so that only models which satisfy this inequality are shown.
As described in the text, a $1\sigma$ region does not appear in the plot because this parameter space does not include any model whose
$\chi^2$ gives better than $1\sigma$ agreement with the WMAP data.}
\label{fig:wmap_ghost}
\end{figure}

The next space of models we consider is the near de Sitter limit with $|d_1| \lesssim 4(d_2+d_3)^{1/2}$ relevant for Ghost inflation.
In this case, we take the model parameters to be $(d_1,d_2+d_3)$.
Combining Eqs.~(\ref{eq:newfnlghost}),~(\ref{eq:relationshipdsghost}) above, the expectation values of the $f_{NL}$ estimators are given by:
\be
\left( \barr{c}
  \langle \hfnlequil(d_1,d_2+d_3) \rangle \\
  \langle \hfnlorthog(d_1,d_2+d_3) \rangle
\earr \right)
=
\left( \barr{cc}
 0.8625 & 0.9685 \\
 0.2621 & 0.1667
\earr \right)
\left( \barr{c}
138.1 (d_2+d_3)^{-4/5} \\
71.97(d_1)(d_2+d_3)^{-13/10}
\earr \right)
\ee
Using this expression, we can define a $\chi^2$ statistic as in Eqs.~(\ref{eq:chisq_near_desitter}) above.
In Fig.~\ref{fig:wmap_ghost}, we show $2\sigma$ and $3\sigma$ contours in the $(d_1,d_2+d_3)$ plane
obtained using this statistic $\chi^2(d_1,d_2+d_3)$.
Note that no $1\sigma$ contour appears in the figure.
This is because there are no models in this parameter space which satisfy 
$\chi^2(d_1,d_2+d_3) \le 2.28$. (This value of $\chi^2$ corresponds to $p$-value $p=0.32$,
and is the threshhold value for the $1\sigma$ region, as described in sec.~\ref{ssec:wmap_single_field}.)
For example, if we try to solve for the values of $d_1,(d_2+d_3)$ which satisfy $\chi^2=0$
(or equivalently, $\langle \hfnlequil \rangle = (\hfnlequil)_{\rm WMAP}$ and 
$\langle \hfnlorthog \rangle = (\hfnlorthog)_{\rm WMAP}$), then we find a negative value
of $(d_2+d_3)^{4/5}$, so no model with $\chi^2=0$ exists.
In contrast, the other parameter spaces considered in this paper do include models with $\chi^2=0$, 
and a $1\sigma$ region appears in the associated figures (Figs.~\ref{fig:csc3},~\ref{fig:wmap_near_desitter}).
The absence of a $1\sigma$ region in Fig.~\ref{fig:wmap_ghost} should not be interpreted as evidence that the near de
Sitter model with $|d_1| \lesssim 4(d_2+d_3)^{1/2}$ is inconsistent with or disfavored by the WMAP data, since it does
include models which are consistent with WMAP at $2\sigma$.

As in the former models, so far we have neglected metric fluctuations. Following again \cite{Creminelli:2006xe,Cheung:2007sv} it is straightforward to see that this is a good approximation for $(d_2+d_3)\lesssim (\mpl/(30 M))^{10}$. For the parameter space we are interested in, this translates in $M\lesssim \mpl/10$ to a good approximation. Similarly to the former case, when this inequality is violated, as shown in  \cite{Creminelli:2006xe,Cheung:2007sv}  the squared speed of sound $c_s^2$ of the fluctuations becomes negative and the modes begin to grow exponentially before crossing the horizon. In this case these models become again similar to the ones shown in the next sec.~\ref{sec:negativec_s^2}, and, as we will see, are generically very disfavored by the data.

\subsection{Near-de-Sitter models with $c_s^2 < 0$ and $-d_1 \gtrsim |4{\dot H} \mpl^2 / (M^3 H)|$}
\label{sec:negativec_s^2}

\begin{figure}
\centerline{ \includegraphics[width=4.1in]{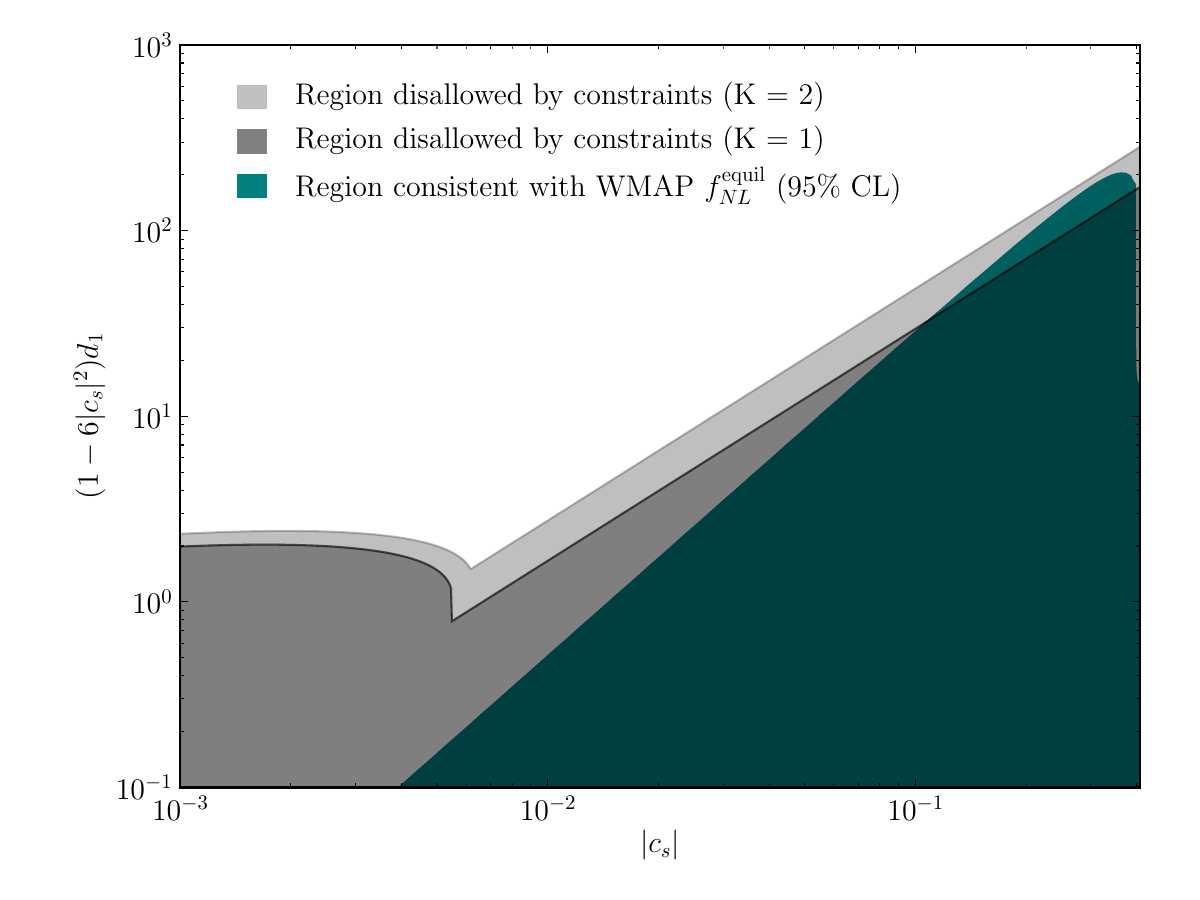} }
\caption{\small Analysis of the near de Sitter model with $c_s^2 < 0$ and $d_1 \lesssim 4{\dot H} \mpl^2 / (M^3 H)$.
We show the region of parameter space consistent with the WMAP upper limit $\fnlequil \lesssim 435$ (95\% CL),
and the parameter constraints that must be satisfied in this model (eq.~(\ref{eq:negcs_case1_constraints})). Notice that $\fnlequil$ must be positive in this model.
Using loose constraints (taking $K=1$ as defined below eq.~(\ref{eq:negcs_case1_constraints})),
a very small region of parameter space is consistent with WMAP.
Using slightly more conservative constraints ($K=2$), the entire parameter space is ruled out at 95\%~ CL.}
\label{fig:wmap_negative_cs2_case1}
\end{figure}

In the near de Sitter model with $c_s^2 < 0$ and $-d_1 \gtrsim |4{\dot H} \mpl^2 / (M^3 H)|$, the bispectrum 
(eq.~(\ref{eq:fnlcs_negative_1})) is always highly correlated to the equilateral shape.
This is because the signs of $f_{NL}^{\dot\pi(\partial\pi)^2}$ and $f_{NL}^{(\partial^2\pi)(\partial\pi)^2}$ are
always the same, so it is not possible to get the near-cancellation which leads to the orthogonal shape.
(More precisely, we find that the cosine between the equilateral shape and the bispectrum in this model is equal to 0.9995 for small $|c_s|$ and becomes even closer to one as we let $|c_s|$ grow.)
For this reason, we will discard $\fnlorthog$ and only use $\fnlequil$ in the analysis.
Using eq.~(\ref{eq:relationshipdscsnegative1}), and neglecting the orthogonal template,
the expectation value of $\hfnlequil$ is given by:
\be
\langle \hfnlequil \rangle = 
\left( \begin{array}{cc}
1.040 & 0.9878
\end{array} \right)
\left( \begin{array}{c}
(2.528 \times 10^{-11}) \frac{d_1^4 (1 - 6|c_s|^2)^4}{|c_s|^7}  \\
(3.867 \times 10^{-11}) \frac{d_1^4 (1 - 6|c_s|^2)^3}{|c_s|^7}
\end{array} \right)
\ee
where here, contrary to what we did in eq.~(\ref{eq:fnlcs_negative_1}), we are keeping the corrections that come from $|c_s|\simeq1$ to be more accurate. The region in the $(|c_s|,d_1)$ plane which is consistent with the WMAP constraint $\fnlequil \lesssim 435$ (95\% CL)
from sec.~\ref{sec:constraints_error} is shown in Fig.~\ref{fig:wmap_negative_cs2_case1}. Notice that in this case $\fnlequil$ has to be positive.

\par From eq.~(\ref{eq:speed_of_sound_negaritve}), the absolute value of the speed of sound is bounded to be
\be
|c_s^2|\leq \frac{1}{6}\ .
\ee
The consistency of the calculation (see \cite{Senatore:2004rj}) forces the model parameters to satisfy the following constraints:
\bea\label{eq:negcs_case1_constraints}
&&\frac{-d_1}{4(d_2+d_3)^{1/2}}\simeq\frac{1}{2} \log\left( \frac{25}{4608} \Delta_\Phi (1 - 6|c_s|^2)^4 \frac{d_1^4}{|c_s|^5} \right)  \gg  1\ , \nn \\
&&\frac{\Lambda}{H}\sim \frac{(d_2+d_3)^{7/2} M}{H}\simeq \frac{(1 - 6 |c_s|^2)(- d_1) (d_2+d_3)^{7/3}}{8 |c_s|^2}\gg 1 \ .\\ \nn
\eea
The first constraint follows from the expression for the power spectrum in
eq.~(\ref{eq:deltacs_negative_1}) and the requirement that $d_1 \lesssim -4(d_2+d_3)^{1/2}$ in this model, which just comes from imposing that the dispersion relation at horizon crossing is dominated by the term is $c_s k$. The second requirement comes from imposing that the cutoff of the model is much larger than Hubble.

To make the constraints precise, let us introduce a ``threshhold'' parameter $K > 1$, and define
the symbol $\gg$ in eq.~(\ref{eq:negcs_case1_constraints}) to mean that the ratio between the left-hand and right-hand sides must be $\ge K$.
As shown in Fig.~\ref{fig:wmap_negative_cs2_case1}, 
if we take $K = 1$ (corresponding to a loose interpretation of the constraints) then a small region in parameter space is consistent with WMAP,
whereas if we take $K = 2$ (corresponding to a slightly more conservative interpretation) then the entire parameter space is ruled out.

As in the former cases, we have neglected metric fluctuations. Following \cite{Creminelli:2006xe,Cheung:2007sv} it is straightforward to see that this is a good approximation for $d_1\lesssim (\mpl/(30 M))^{5}$. When this inequality is violated, as shown in  \cite{Creminelli:2006xe,Cheung:2007sv}  the squared speed of sound $c_s^2$ of the fluctuations is still negative. In this case the model is still very similar to the one just shown, and, as we are seeing, it is generically very disfavored by the data.

\subsection{Near-de-Sitter models with $c_s^2 < 0$ and $|d_1| \lesssim 4{\dot H} \mpl^2 / (M^3 H)$}

In this last case, the bispectrum is given by:
\be
\left( \begin{array}{c}
  \langle \hfnlequil \rangle  \\
  \langle \hfnlorthog \rangle
\end{array} \right)
=
\left( \begin{array}{cc}
  1.040 & 0.9878 \\
  0.1079 & 0.1613
\end{array} \right)
\left( \begin{array}{c}
  (5.179 \times 10^{-8}) \left( \frac{{\dot H} \mpl^2 (1+|c_s|^2)}{H^4 |c_s|} \right) \\
  (-1.177 \times 10^{-8}) \left( \frac{{\dot H} \mpl^2 (1+|c_s|^2)}{H^4 |c_s|} \right)^{3/4} \frac{d_1}{|c_s|^{7/4}}
\end{array} \right)  \label{eq:estimators_negative_cs_case2}
\ee
by combining Eqs.~(\ref{eq:fnlcs_negative_2}) and~(\ref{eq:relationshipdscsnegative1}). Notice that in this model $\dot H>0$ and that, contrary to what we did in eq.~(\ref{eq:fnlcs_negative_1}), we are keeping the corrections that come from $|c_s|\simeq1$ to be more accurate.

\begin{figure}
\centerline{ \includegraphics[width=3.8in]{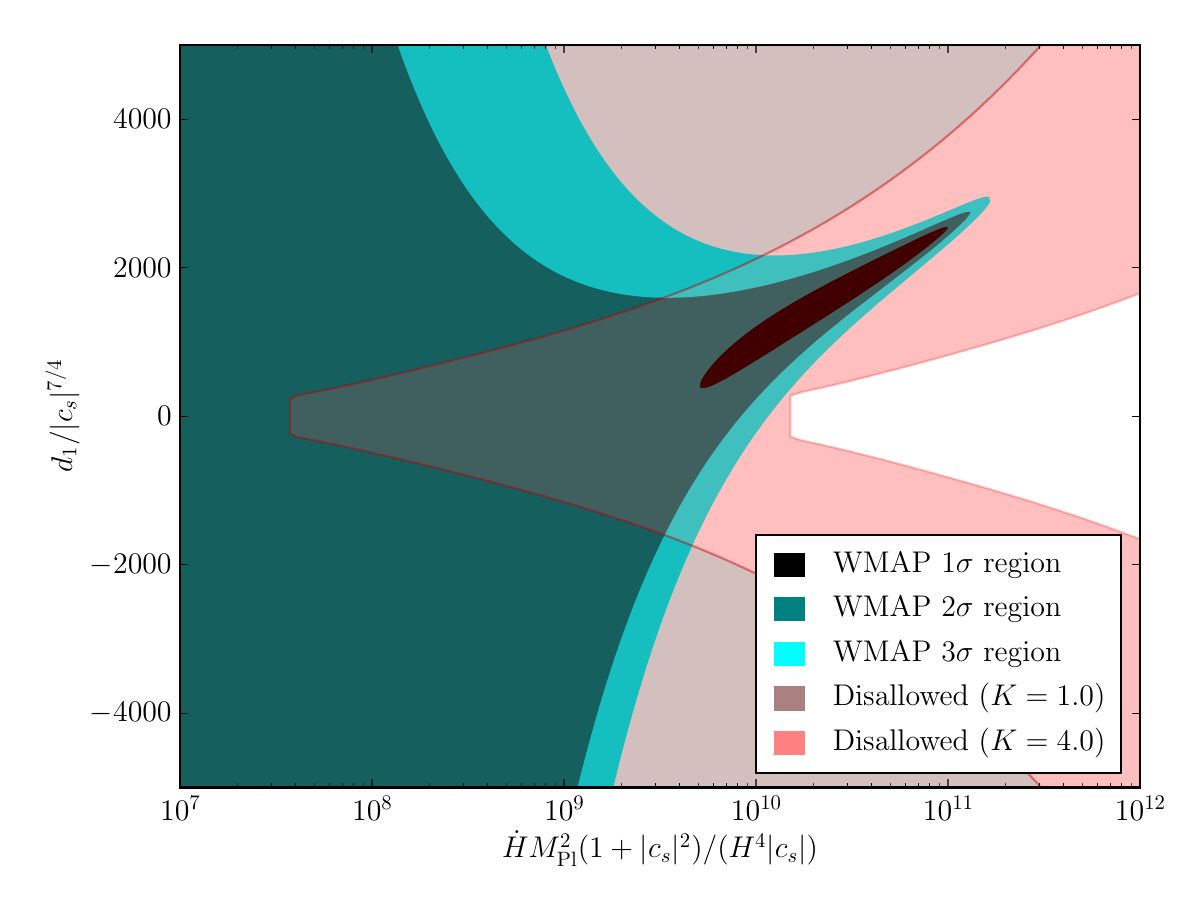} }
\caption{\small Analysis of the near de Sitter model with $c_s^2 < 0$ and $|d_1| \lesssim 4{\dot H} \mpl^2 / (M^3 H)$. Notice that this model has $\dot H>0$.
We show the regions consistent with WMAP, defined using a $\chi^2$ statistic which includes $\fnlequil$ and $\fnlorthog$,
and the regions that are disallowed by the constraints in eq.~(\ref{eq:negcs_case2_constraints}).
Using loose constraints (taking $K=1$ as defined below eq.~(\ref{eq:negcs_case2_constraints})), there is a large region
of parameter space which is consistent with WMAP, including the ``branch'' in the upper right part of the plot where
$\fnlorthog$ is generated.  Using somewhat more conservative constraints ($K=4$), the entire parameter space is ruled out.}
\label{fig:wmap_negative_cs2_case2}
\end{figure}

Because the bispectrum depends on two combinations of parameters, we define a $\chi^2$ statistic which
uses both $\fnlequil$ and $\fnlorthog$, by using eq.~(\ref{eq:chisq_near_desitter}) with expectation values given by
eq.~(\ref{eq:estimators_negative_cs_case2}).
In Fig.~\ref{fig:wmap_negative_cs2_case2}, we show the region of parameter space which is consistent with WMAP,
using this statistic.

We have the following parameter constraints in this model:
\bea \label{eq:negcs_case2_constraints}
&&\frac{|d_1|}{|c_s|^{7/4}}  \ll  \frac{2^{11/4}}{(1+|c_s|^2)} \left( \frac{{\dot H} \mpl^2 (1+|c_s|^2)}{H^4 |c_s|} \right)^{1/4}  \ , \\ \nn
&& \frac{2 |c_s|^2  M}{(d_2+d_3)^{1/2} H}=\frac{1}{2}\log\left(\frac{100}{9} \Delta_\Phi|c_s|(1+|c_s^2|)\frac{\dot H \mpl^2}{H^4}\right)\gg 1\ , \\ \nn
&& \frac{\Lambda}{H}\sim \frac{(d_2+d_3)^{7/2} M}{H}\simeq \frac{128\cdot 2^{11/12}\cdot |c_s|^{13/2}(1+|c_s|^2)^{17/12}}{\log\left(\frac{100}{9} \Delta_\Phi|c_s|(1+|c_s^2|)\frac{\dot H \mpl^2}{H^4}\right)^{14/3}}\left(\frac{\dot H \mpl^2}{H^4}\right)^{17/12}\gg 1 \ .
\eea
The first constraint follows from the requirement that $|d_1| \lesssim (4 {\dot H} \mpl^2) / (M^3 H)$ in this model, together with
the expression for the sound speed in eq.~(\ref{eq:cs_negative_2}). The second constraint comes from imposing that the dispersion relation at horizon crossing is dominated by the term is $c_s k$, while the third comes from imposing the cutoff scale to be larger than $H$.
As in the previous subsection, we define the symbol $\ll$ in eq.~(\ref{eq:negcs_case2_constraints})
by introducing a threshhold parameter $K$: we say that $x \ll y$ if the ratio $(y/x)$ is $\ge K$.

If we use loose constraints $(K=1)$, then there is a large region of parameter space which is consistent with WMAP,
including an extended ``branch'' which points toward the upper right corner in Fig.~\ref{fig:wmap_negative_cs2_case2}.
This branch corresponds to models in which the contributions to the bispectrum from the operators $\dot\pi(\partial\pi)^2$ 
and $(\partial^2\pi)(\partial\pi)^2$ have opposite signs and nearly cancel, so that the orthogonal shape is generated instead
of the equilateral shape.
Because our analysis includes $\fnlorthog$, the branch is cut off and arbitrarily large values of the parameters are not allowed.

If we use somewhat more conservative constraints $(K=4)$, then the entire parameter space is ruled out, including the branch.

As in the former cases, we have neglected metric fluctuations. Following \cite{Creminelli:2006xe,Cheung:2007sv} it is straightforward to see that this is a good approximation for $d_2+d_3\lesssim (\mpl/(30 M))^{10}$. When this inequality is violated, as shown in  \cite{Creminelli:2006xe,Cheung:2007sv}  the squared speed of sound $c_s^2$ of the fluctuations is still negative. In this case the model is still very similar to the one just shown, and, as we are seeing, it is generically very disfavored by the data.

\section{Summary\label{sec:summary}}


The non-Gaussianities produced by the most general single-field inflation with an approximate shift symmetry protecting the Goldstone boson are described by two independent parameters. The resulting shape of the signal in Fourier space can vary from being peaked on equilateral triangles to being peaked on flat triangles (where the two shortest sides are equal to half the longest one). Two independent parameters control this bi-dimensional space of signatures and they both need to be constrained. We have shown that former analyses have been partially blind to a fraction of this parameter space.

We have then shown that an analysis covering  the whole of the parameter space can be performed in practice using two approximate factorizable templates. The first template is the standard equilateral one, which constrains the amount of signal that is peaked on equilateral configurations in Fourier space. We call the second template `orthogonal'  and we choose it to constrain the amount of signal that in Fourier space is peaked both on equilateral configurations and on flat triangular configurations (with opposite signs). This new shape is approximately orthogonal to the equilateral one using a 3D scalar product.

We apply the optimal estimator for the two $f_{NL}$ parameters that characterize the amplitude of the two independent templates to the WMAP 5-year data. We find no evidence of non-Gaussianities, and we obtain the following optimal constraints:
\bea
&&-125\le \fnlequil\le 435  \quad {\rm at \ 95\%\ \ CL}\ , \\
&&-369\le \fnlorthog\le 71 \quad {\rm at \ 95\%\ \ CL}\ .
\eea
The $\mbox{Corr}(\hfnlequil, \hfnlorthog)$  in the Monte Carlo simulations is equal to $0.32$.
The constraint on $\fnlequil$ is milder than the one presented by the WMAP collaboration for the same set of data \cite{Komatsu:2008hk}. We believe that previous analyses
have underestimated the error bar on $f_{NL}^{\rm equil}$ and that our constraints
are optimal.

Thanks to the effective field theory of inflation \cite{Cheung:2007st}, we are able to map the constraints on the two $f_{NL}$ parameters above into constraints on the coefficient of the interaction Lagrangian of the Goldstone boson. Under the assumption that the primordial density perturbations are generated by a single-field inflationary model where there is an approximate shift symmetry for the Goldstone boson, this mapping is unique and constrains all possible inflationary models of this kind. We stress again that we do not assume that the background solution is given by a fundamental scalar field: the Lagrangian for the fluctuations in terms of the Goldstone boson is independent of the details through which the background solution is generated. We are only assuming that there is one light degree of freedom playing a relevant role during the inflationary phase. We draw contour plots for the parameters of the Goldstone Lagrangian that are constrained by our analysis. In particular, for one of them, the speed of sound of the inflaton fluctuations, we find that it has to be larger than:
\be
c_s\ge 0.011 \qquad {\rm at \ 95\%\ \ CL}\ ,
\ee
or smaller than
\be
c_s\lesssim 10^{-2}\;(d_2+d_3)^{2/5}\ .
\ee
In this last case the higher-derivative kinetic term is important at horizon crossing and the non-Gaussianities depend on other coefficients. In some region of the parameter space, consistent inflationary models have a negative squared speed of sound $c_s^2$ for the fluctuations at horizon crossing. This leads to an exponential growth of the perturbations before horizon crossing and an associated increase in the level of non-Gaussianities. The analysis of the WMAP data shows that these models are practically ruled out at 95\% CL.

These error bars are expected to decrease by a factor of six in the next few years when the data from the Planck satellite will be available \cite{Babich:2004yc}.

\section*{Acknowledgments}
We would like to thank Eiichiro Komatsu for help during the project and in particular for some comparisons of our code with the one used by the WMAP team. We thank Nima Arkani-Hamed, Shamit Kachru, Uros Seljak, Eva Silverstein, David Spergel and Jay Wacker for useful discussions.  
LS was supported in part by the National Science Foundation under Grant No. PHY-0503584. KMS was supported by an STFC Postdoctoral Fellowship.
 MZ was supported by NASA NNG05GJ40G and NSF AST-0506556 as well as the David and Lucile Packard, Alfred P. Sloan and John D. and Catherine T. MacArthur foundations. 
KMS would like to thank the hospitality of the Department of Astrophysics at Princeton University, where this work was partially carried out.

 \appendix

\section{A different but equivalent definition of the estimator\label{app:other_definition}}
 
 In the main text, we related  $\fnlequil$ and $\fnlorthog$ to $c_s$ and $\tilde c_3$ in eq.~(\ref{eq:relationshipp}) by noticing that the expectation value of our estimator was given by the 3D Fisher matrix between the shape of non-Gaussianity in single field inflation and the templates. If instead one wanted to define $\fnlequil(c_s,\tilde c_3)$ and $\fnlorthog(c_s,\tilde c_3)$ so that our template shapes best approximate the exact expression for given values of Lagrangian parameters $c_s$ and $\tilde c_3$, then one would instead use  the following expression~\footnote{Here, for simplicity, we will refer only to the case where inflation happens in the way that in the text we are referring to as 'not-so-near-de-Sitter'. The discussion applies with trivial modifications to when inflation happens near-de-Sitter. }:
 \bea\label{eq:relationship_second}
&&\left( \barr{c}
  \fnlequil(c_s,\tilde c_3)  \\
   \fnlorthog(c_s,\tilde c_3) 
\earr \right)=\\ \nonumber
&&=\left( \barr{cc}
  F_{\rm equil.}{}_{,\,1} \cdot F_{\rm equil.}{}_{,\,1} &   F_{\rm equil.}{}_{,\,1} \cdot F_{\rm orthog.}{}_{,\,1}  \\
   F_{\rm equil.}{}_{,\,1} \cdot F_{\rm orthog.}{}_{,\,1} &   F_{\rm orthog.}{}_{,\,1} \cdot F_{\rm orthog.}{}_{,\,1}
\earr \right)^{-1}\cdot \left( \barr{c}
   F_{\rm equil.}{}_{,\,1} \cdot F \\
   F_{\rm orthog.}{}_{,\,1} \cdot F
\earr \right)\\ \nonumber
&&= \left( \barr{cc}
 1.056 & 1.283 \\
 -0.05721 & -0.2663
\earr \right)
\left( \barr{c}
 f_{NL}^{\dot\pi(\d_i\pi)^2}(c_s)   \\
  f_{NL}^{\dot\pi^3}(c_s,\tilde c_3) \earr \right)\ ,
\eea
where the subscript $_1$ means that the shape is evaluated with $f_{NL}=1$.
 This expression is in fact obtained by minimizing the Fisher Matrix distance defined as the norm of the shape $F-F_{\rm template}$ according to the 3D scalar product.
 
The reason why in the main text we use instead the definition in~(\ref{eq:relationshipp}) is that, for what  the data analysis is concerned, we are interested in the $\chi^2$ statistics, and this depends on the definition of the estimator. In fact, expression~(\ref{eq:relationshipp}) represents the expectation value of the estimator in the presence of a given ${f}_{NL}^{\rm\dot \pi(\d_i\pi)^2}$ and ${f}_{NL}^{\rm \dot\pi^3}$, in the approximation that the 3D scalar product is a good approximation to the 2D one.
This is the correct relationship according to the definition of the estimator we use and that we explain in the main text, where we analyze the data for the equilateral and the orthogonal shape assuming, in each case, that the other shape gives zero contribution.

However, this is not the only definition of the estimator one could have used. One could have in fact imagined to build an estimator for each shape that has zero response to the other shape. This different definition {\it would} give different results for the constraints on $\fnlequil$ and $\fnlorthog$, but, as we are now going to show, it {\it would not} change the $\chi^2$ statistics from which we deduce the constraints on the parameters of the Lagrangian.
Let us see in detail how this happens, and to this purpose let us define the matrix $F$ of the scalar products between the templates as
\bea
F=\left(\barr{cc}
(F_{\rm equil.}{}_{,\,1}\cdot F_{\rm equil.}{}_{,\,1}) & (F_{\rm equil.}{}_{,\,1}\cdot F_{\rm orthog.}{}_{,\,1}) \\
(F_{\rm equil.}{}_{,\,1}\cdot F_{\rm orthog.}{}_{,\,1}) & (F_{\rm orthog.}{}_{,\,1}\cdot F_{\rm orthog.}{}_{,\,1})
\earr
\right)\ ,
\eea
 and the matrix $M$ of the scalar products between the templates and the single field shapes as
\bea\label{eq:matrixM}
M=\left(\barr{cc}
(F_{\rm equil.}{}_{,\,1}\cdot F_{\rm \dot\pi(\d_i\pi)^2}) & (F_{\rm equil.}{}_{,\,1}\cdot F_{\rm \dot\pi^3}) \\
(F_{\rm orhtog.}{}_{,\,1}\cdot F_{\rm \dot\pi(\d_i\pi)^2}) & (F_{\rm orthog.}{}_{,\,1}\cdot F_{\rm \dot\pi^3})
\earr
\right)\ .
\eea
In the two equations above, as well as in the rest of this Appendix, the dot $\cdot$ is now meant to represent a 2D scalar product as defined in eq.~(\ref{eq:2D_scalar_product}). For simplicity, we call the induced byspectrum $B_{i}$ by a shape $F_i$ simply as $F_i$.

Let us also define the vector of the two estimators
\bea
\hat E=
\left( \barr{c}
\hat{\tilde f}_{NL}^{\rm equil.} \\
\hat{\tilde f}_{NL}^{\rm orthog.}
\earr \right)\ ,
\eea 
where the $\tilde {\hat f}$ represents the fact that we have removed the normalization factors $N_{\rm equil.}=F_{11}$ and $N_{\rm orthog.}=F_{22}$ from eq.~(\ref{eq:equil_estimator}) and (\ref{eq:orthog_estimator})   respectively. We also define the vector $f$ of the $f_{NL}$'s as
\bea
f=
\left( \barr{c}
{f}_{NL}^{\rm\dot \pi(\d_i\pi)^2} \\
{f}_{NL}^{\rm \dot\pi^3}
\earr \right)\ .
\eea 

The way we define the estimator $\hat E^{(1)}$ that we use in in our analysis is by normalizing the estimator $\hat E$ in the following  way:
\bea
\hat E^{(1)}=
\left( \barr{c}
\frac{1}{F_{11}}\hat{\tilde f}_{NL}^{\rm equil.} \\
\frac{1}{F_{22}}\hat{\tilde f}_{NL}^{\rm orthog.}
\earr \right)\ ,
\eea 
Since at the moment we are not seeing any evidence for a non-zero value for any of the $f_{NL}$'s, the estimator $E^{(1)}$ analyzes the data for each shape assuming that the other shape is not present. The covariance $C^{(1)}$ for this estimator is given by the matrix
\bea
C^{(1)}=\left(\barr{cc}
\frac{1}{F_{11}}& \frac{F_{12}}{F_{11}F_{22}} \\
\frac{F_{12}}{F_{11}F_{22}}  & \frac{1}{F_{22}}
\earr
\right)\ ,
\eea
In the case of the estimator $E^{(1)}$, it is straightforward to see that the $\chi{}^2$ statistics is given by
\be
\chi_{(1)}^2(f)=(f-f_{\rm WMAP})^T M^T F^{-1} M (f-f_{\rm WMAP})\ ,
\ee
where $f_{\rm WMAP}$ is the result of $f$ deduced from the WMAP data, and we have made use of the matrix $M$ to translate the values of $\fnlequil$ and $\fnlorthog$ into values for  ${f}_{NL}^{\rm\dot \pi(\d_i\pi)^2}$ and ${f}_{NL}^{\rm \dot\pi^3}$. The matrix $M$ we use in the analysis is analogous to the one defined in eq.~(\ref{eq:matrixM}), with the 2D scalar product replaced with 3D scalar products. We believe that this is a reasonable approximation, at order $10\%$ level, given that the templates we use are very similar to the actual shape. This level of accuracy is in particular justified by the current obervational status, where we do not have a detection, but only constraints. 

Now, we could instead have defined for our analysis the estimator in an alternative way, that we call $\hat E^{(2)}$:
\be
\hat E^{(2)}=F^{-1} E\ .
\ee 
This definition is also commonly used in the literature.
With this definition, there is no cross-response between the equilateral and the orthogonal shape, and therefore the expectation value of the estimator in the presence of an hypothetical non-Gaussianity in the sky induced by single field inflation with a given $c_s$ and $\tilde c_3$, is given by eq.~(\ref{eq:relationship_second}) with the 3D scalar products replaced with 2D ones. In this case, the estimator covariance $C^{(2)}$ is simply given by
\be
C^{(2)}=F^{-1}\ ,
\ee
and the $\chi{}^2$ is given by
\bea
\chi_{(2)}^2(f)&= &(f-f_{\rm WMAP})^T M^T F^{-1} F^{-1} F M (f-f_{\rm WMAP})=\\ \nonumber
&&(f-f_{\rm WMAP})^T M^T F^{-1} M (f-f_{\rm WMAP})\ .
\eea
where we have used that the matrix in eq.~(\ref{eq:relationship_second}) is equal to $F^{-1}M$ and we have made the same approximation as before in approximating 2D scalar products with 3D scalar  products.
The $\chi^2$ defined for the two estimators $E^{(1)}$ and $E^{(2)}$, using respectively the mappings of eq.~(\ref{eq:relationshipp}) and eq.~(\ref{eq:relationship_second}), are equivalent. This means that while the constraints we get on $\fnlequil$ and $\fnlorthog$ {\it do} depend on the choice of the estimator (and we explain in the main part of the text why we prefer the first definition of the estimators), the constraints on the parameters of the Lagrangian, {\it do not} depend on these two choices. 
A very similar argument to the one above shows that our constraints on single-field inflation would be unchanged if we used a different value of $c$ to define $\fnlorthog$ in eq.~(\ref{eq:generic_Forthog}).

\section{A more accurate template\label{App:template}}
 
 We have found that a different choice of templates with respect to the one we make in the main text allows for a better approximation of the bi-dimensional space of non-Gaussianities produced in single field inflation.  This is achieved if
one substitutes the orthogonal template we have used in the main part of the paper with the following one, which is again similar to the orthogonal shape (see upper-right panel of Fig.~\ref{fig:shape}):
\bea
&&F_{\rm orthog.\, (2)} = f_{NL}^{\rm orthog}\cdot6 \Delta_\Phi^2
   \left[ (1+p) \frac{\Delta(k_1,k_2,k_3)}{k_1^3 k_2^3 k_3^3}
      - p \frac{\Gamma(k_1,k_2,k_3)^3}{k_1^4 k_2^4 k_3^4}\right]\ .
\eea
Here $p=\frac{27}{-21+\frac{743}{7(20\pi^2-193)}}\simeq 8.52\,$, and
\be
\Delta(k_1,k_2,k_3)=(k_t-2 k_1)(k_t-2 k_2)(k_t-2 k_3)\ ,
\ee
with $k_t=k_1+k_2+k_3$, and
\be
\Gamma(k_1,k_2,k_3)=\frac{2}{3}(k_1 k_2+k_2 k_3+ k_3 k_1)-\frac{1}{3}(k_1^2+k_2^2+k_3^2)\ .
\ee
The value of $p$ is chosen in such a way that the template $F_{\rm orthog. \ (2)}$ has the useful property of being orthogonal to the template $F_{\rm equil}$, in the sense of the 3D scalar product between shapes of eq.~(\ref{eq:scalar_product}). In practice, $F_{\rm equil}$ is a good template for the equilateral shape with $\tilde c_3\simeq 0$, while  $F_{\rm orthog.}$  is a good template for the orthogonal shape $\tilde c_3\simeq -5.4$.

Similarly to what we did in the main part of the text, we define the template for single-field inflation to be:  
\be
F_{\rm singl.\ field\, (2)}(k_1,k_2,k_3,c_s,\tilde c_3)= f_{NL}^{\rm equil.}(c_s,\tilde c_3)F_{\rm equil.}(k_1,k_2,k_3)+f_{NL}^{\rm orthog. \,(2)}(c_s,\tilde c_3)F_{\rm orthog.\, (2)}(k_1,k_2,k_3)\ .
\ee
The analogous of eq.~(\ref{eq:relationshipp}), now becomes
\bea\label{eq:relationship}
\left( \barr{c}
  \fnlequil(c_s,\tilde c_3)  \\
   \fnlorthog(c_s,\tilde c_3) 
\earr \right)
&=& 
\left( \barr{cc}
  \left( \frac{F_{\dot\pi (\partial\pi)^2} \cdot F_{\rm equil.}}{F_{\rm equil.} \cdot F_{\rm equil.}} \right) &
     \left( \frac{F_{\dot\pi^3} \cdot F_{\rm equil.}}{F_{\rm equil.} \cdot F_{\rm equil.}} \right) \\
  \left( \frac{F_{\dot\pi (\partial\pi)^2} \cdot F_{\rm orthog.\, (2)}}{F_{\rm orthog.\, (2)} \cdot F_{\rm orthog.\, (2)}} \right) &
     \left( \frac{F_{\dot\pi^3} \cdot F_{\rm orthog.\, (2)}}{F_{\rm orthog.\, (2)} \cdot F_{\rm orthog.\, (2)}} \right)
\earr \right)_{f_{\rm NL}\cdot \Delta_\Phi^2=1}
\left( \barr{c}f_{NL}^{\dot\pi(\d_i\pi)^2}(c_s) \\ 
 f_{NL}^{\dot\pi^3}(c_s,\tilde c_3) 
\earr \right)  \nn \\
&=& \left( \barr{cc}
 1.040 & 1.210 \\
 -0.03951 & -0.1757
\earr \right)
\left( \barr{c}
 f_{NL}^{\dot\pi(\d_i\pi)^2}(c_s)   \\
  f_{NL}^{\dot\pi^3}(c_s,\tilde c_3) \earr \right)
\eea
where $ f_{NL}^{\dot\pi(\d_i\pi)^2}(c_s)$ and  $ f_{NL}^{\dot\pi(\d_i\pi)^2}(c_s,\tilde c_3)$ are given in eq.~(\ref{eq:newfnl}).
 
Quite remarkably, the scalar product of this template with the exact shape (in the case not-near-de-Sitter) for any value of $c_s$ and $\tilde c_3$ is always larger than $0.99$ (the minimum is at $\tilde c_3\simeq -5.4$). This tells us that these two independent templates are effectively covering all the parameter space of shapes generated in single-field inflation. The reason why this second orthogonal template does such a better job than the one we use in the text is that in the squeezed limit $k_1\ll k_2,k_3$, $F_{\rm orthog.\,(2)}(k_1,k_2,k_3)\sim{\cal{O}}(1/k_1)$ as the exact shape, while the template we use in the text goes as ${\cal{O}}(1/k_1^2)$. The amount of signal contained in this limit is not very large, and this is why the orthogonal template we use in the main text gives still a good approximation to the exact shape. 
We have found that implementing this template numerically is quite difficult, and therefore, given than the orthogonal template we present in the main text does not create such numerical difficulties and it is still accurate enough, we have decided to perform the analysis of the data with the $F_{\rm orthog.}$ presented in the text. 
   
 \begingroup\raggedright\endgroup

\end{document}